\documentclass[
 aip,
 amsmath,amssymb,
 reprint,
floatfix
]{revtex4-1}
\usepackage{graphicx}
\usepackage{dcolumn}
\usepackage{bm}
\usepackage{tikz}
\usepackage{siunitx}
\usepackage{multirow}
\usepackage{amsmath}
\DeclareMathOperator*{\argmax}{arg\,max}
\DeclareMathOperator{\spn}{span}

\usepackage[utf8]{inputenc}
\usepackage[T1]{fontenc}
\usepackage{mathptmx}
\usepackage{etoolbox}
\usepackage{lipsum}
\usepackage[hidelinks]{hyperref}

\newcommand{\ip}[3]{\left\langle#1, #2\right\rangle_{#3}}
\newcommand{\mean}[3]{\ensuremath{\left\langle\left\{#1\right\}_{#2}^{#3}\right\rangle}}

\makeatletter
\def\@email#1#2{%
 \endgroup
 \patchcmd{\titleblock@produce}
  {\frontmatter@RRAPformat}
  {\frontmatter@RRAPformat{\produce@RRAP{*#1\href{mailto:#2}{#2}}}\frontmatter@RRAPformat}
  {}{}
}%
\makeatother
\begin{document}

\preprint{AIP/123-QED}

\title{Dissipation-based proper orthogonal decomposition of turbulent Rayleigh-Bénard convection flow}
\author{P.J.~Olesen}
\email{pjool@dtu.dk}
\affiliation{ 
Department of Civil and Mechanical Engineering, 
Technical University of Denmark, 
Koppels Allé 403, 
2800 Kgs.~Lyngby, 
Denmark
}
\author{L.~Soucasse}
\affiliation{
Université Paris-Saclay, CNRS, CentraleSupélec, Laboratoire EM2C, 91190 Gif-sur-Yvette, France.}
\affiliation{Netherlands eScience Center,
Science Park 402 (Matrix III),
1098 XH Amsterdam,
Netherlands}
\author{B.~Podvin}
\affiliation{
Université Paris-Saclay, CNRS, CentraleSupélec, Laboratoire EM2C, 91190 Gif-sur-Yvette, France.}
\author{C.M.~Velte}
\affiliation{ 
Department of Civil and Mechanical Engineering, 
Technical University of Denmark, 
Koppels Allé 403, 
2800 Kgs.\ Lyngby,
Denmark
}

\date{\today}
\begin{abstract}
We present a formulation of proper orthogonal decomposition (POD) producing a velocity-temperature basis optimized with respect to an $H^1$ dissipation norm. This decomposition is applied, along with a conventional POD optimized with respect to an $L^2$ energy norm, to a data set generated from a direct numerical simulation of Rayleigh-Bénard convection in a cubic cell ($\mathrm{Ra}=10^7$, $\mathrm{Pr}=0.707$). The data set is enriched using symmetries of the cell, and we formally link symmetrization to degeneracies and to the separation of the POD bases into subspaces with distinct symmetries. We compare the two decompositions, demonstrating that each of the 20 lowest dissipation modes is analogous to one of the 20 lowest energy modes. Reordering of modes between the decompositions is limited, although a corner mode known to be crucial for reorientations of the large-scale circulation is promoted in the dissipation decomposition, indicating suitability of the dissipation decomposition for capturing dynamically important structures. Dissipation modes are shown to exhibit enhanced activity in boundary layers. Reconstructing kinetic and thermal energy, viscous and thermal dissipation, and convective heat flux, we show that the dissipation decomposition improves overall convergence of each quantity in the boundary layer. Asymptotic convergence rates are nearly constant among the quantities reconstructed globally using the dissipation decomposition, indicating that a range of dynamically relevant scales are efficiently captured. We discuss the implications of the findings for using the dissipation decomposition in modeling, and argue that the $H^1$ norm allows for a better modal representation of the flow dynamics.
\end{abstract}

\maketitle

\section{Introduction\label{sec:introduction}}

Flows driven by thermal buoyancy are both of fundamental interest and of relevance to a wide range of applications within e.g.~engineering, geophysics, and astrophysics. Such flows are often studied through simplified configurations such as the Rayleigh-Bénard cell, in which a fluid layer between two horizontal plates is heated from the bottom and cooled from the top. For a given geometry the relevant control  parameters are the Rayleigh number, $\mathrm{Ra}$, and the Prandtl number, $\mathrm{Pr}$,
\begin{align}
    \mathrm{Ra} &= \frac{\beta g \Delta T H^3}{\nu \kappa}\,, & \mathrm{Pr} &= \frac{\nu}{\kappa}\,,
    \label{eq:rayleigh}
\end{align}
where $\beta$ is the thermal expansion coefficient, $g$ the acceleration due to gravity, $\Delta T$ the temperature difference between the plates, $H$ the fluid layer height, $\nu$  the kinematic viscosity, and $\kappa$ the thermal diffusivity. The Rayleigh number measures the ratio of buoyancy and dissipative effects, while the Prandtl number depends only on the fluid, measuring the ratio of momentum diffusivity to thermal diffusivity. The resulting behavior of the flow is described through the Nusselt number, $\mathrm{Nu}$, and the Reynolds number, $\mathrm{Re}$:
\begin{align}
    \mathrm{Nu} &= \frac{\left\langle w T\right\rangle_{A,t} - \kappa \left\langle\partial_z T\right\rangle_{A,t}}{\kappa \Delta T H^{-1}}\,,& \mathrm{Re} &= \frac{UH}{\nu}\,,
\end{align}
where $\left\langle\cdot\right\rangle_{A,t}$ denotes the average over a horizontal plane and time, $w$ is the vertical velocity component, $T$ the temperature, $\partial_z$ the vertical derivaive, and $U$ a characteristic mean velocity.

The Nusselt number is the non-dimensionalized heat flux, including both convective and conductive contributions. A central question is to determine the dependence of the heat transfer, characterized by a Nusselt number, on the control parameters $\mathrm{Ra}$ and $\mathrm{Pr}$. \citet{shraiman1990heat} have established strong relationships between these parameters and the thermal and viscous dissipation, valid in the case of adiabatic side walls, showing in particular that \hbox{$\mathrm{Nu} = \left\langle \epsilon_{\theta}\right\rangle$}, where $\left\langle\epsilon_{\theta}\right\rangle$ is the non-dimensional volume-averaged total thermal dissipation, and that $\left(\mathrm{Nu}-1\right)\mathrm{Ra}= \left\langle\epsilon_u\right\rangle$, where $\left\langle\epsilon_{u}\right\rangle$ is the non-dimensional volume-averaged total viscous dissipation.

\citet{grossmann2000scaling} have developed a successful theory for predicting $\mathrm{Nu}$ from $\mathrm{Ra}$ and $\mathrm{Pr}$. It is based on a scaling analysis of the relative contributions of the bulk and the boundary layer to the global kinetic and thermal dissipation, taking into account the relative thickness of the thermal and viscous boundary layers. Crucial dynamical information is therefore contained in these boundary layers,  as pointed out by \citet{siggia1994high}, \citet{chilla2012new}, and more recently by \citet{scheel2014local}, who identified strong inhomogeneities in the boundary layers using  local analysis. At the same time, boundary layers present a challenge in the modeling of buoyancy-driven flows, as pointed out by \citet{hanjalic2002one}, due to rapidly varying properties and the lack of universal scalings for these regions. The theory was further refined by \citet{grossmann2004fluctuations} to decompose the thermal bulk contribution into a background part and a plume-dominated one. Plumes are coherent structures, akin to detached thermal boundary layers, that carry a large part of the  heat flux \citep{shang2003measured} and that are associated with large thermal dissipation events \citep{emran2012conditional}. Plume-based analysis was recently used by \citet{vishnu2022statistics} to investigate the statistics of viscous and thermal dissipation rates in a cubic cell. \citet{shishkina2008analysis} found that plumes collect into a large-scale circulation (LSC) at a sufficiently high Rayleigh number. The connection between small-scale plumes and  the onset of the LSC has also been evidenced in experiments\citep{xi2004laminar}. The presence of the LSC is responsible for the failure of standard one-point turbulence closures \cite{hanjalic2022modelling}.

Coherent pattern extraction from velocity and thermal fluctuations can be carried out using proper orthogonal decomposition (POD)\citep{berkooz1993proper,berkooz1996turbulence}, a statistical technique introduced to turbulence studies by \citet{lumley1967structure}. The first such study of a convective flow was carried out by \citet{sirovich1990turbulent}. Several configurations have been studied since: \citet{bailon2010aspect} applied POD to a cylindrical configuration; \citet{podvin2012proper, verdoold2009prime} to rectangular cavities; \citet{podvin2015large} to two-dimensional cells; and \citet{soucasse2019proper} to cubic cells. In all these studies, the decomposition extracts the velocity and temperature fluctuations with the largest energy. However, as described above, evidence suggests that the viscous and the thermal dissipation rates are the fundamental quantities to characterize the flow. This relates to a more general issue affecting POD-based models, namely that convergence in energy does not itself guarantee that the model reproduces temporal dynamics accurately. In particular, the focus on large-scale structures resulting from energy optimization contrasts with the fundamentally multi-scale nature of turbulence dynamics\citep{bergmann2009enablers}. 
Convergence of the gradients ($H^1$ norm) is a stronger requirement than $L^2$ convergence, and may be necessary for ensuring that the dynamics of the reduced-order system reproduce those of the full flow. \citet{aubry1993preserving} showed that a truncation containing more than 99\% of the energy was not enough to capture the dynamics of the Kuramoto-Sivashinsky equation. It is therefore of interest to investigate the ability of POD-based methods to capture structures optimized with respect to $H^1$ norms rather than the conventional $L^2$ norms.

The formalism underlying POD can be modified to produce decompositions with respect to any quantity that can be expressed formally as a norm on some tensor-, vector-, or scalar-valued field derived from the decomposed data. Using POD this way produces a basis spanning the space corresponding to the field in question optimally with respect to the chosen norm. Examples of such norms are enstrophy and dissipation, which are $H^1$ norms computed from the velocity gradient field and strain rate tensor field, respectively. \citet{sengupta2004proper} studied the bypass transition using an enstrophy-optimized vorticity decomposition, and \citet{lee2020improving} investigated a gradient basis used in combination with a conventional energy-optimized velocity basis as a method for stabilizing a reduced order model (ROM) of a two-dimensional lid driven cavity flow. Another example of POD applied to non-conventional data sets was presented by \citet{schiodt2022characterizing}, who decomposed Lagrangian velocity data of particles suspended in a turbulent flow. 

The extended POD introduced by \citet{boree2003extended} makes it possible to select the decomposed quantity and the optimization quantity independently, enabling the computation of modes spanning any desired quantity optimally with respect to a norm computed from a different quantity, subject to the constraints discussed above. For example, it allows the computation of velocity modes optimized with respect to temperature norm and vice versa, as done by \citet{podvin2015large}, or of velocity modes optimized with respect to the dissipation norm, as demonstrated by \citet{olesen2023dissipation}. This provides a method for educing dynamically important structures that may be missed in a conventional POD analysis. In the present work this approach is used for obtaining velocity-temperature modes that are optimized with respect to either total energy or total dissipation. In the light of the role of dissipation in Rayleigh-Bénard convection discussed above, the dissipation optimization can be expected to produce structures representing important aspects of the flow dynamics that are not well captured in energy optimized decompositions.

In the present work we consider a cubic Rayleigh-Bénard cell, which has previously been investigated with standard POD by \citet{soucasse2019proper}. We apply the dissipation optimized POD to the data set, and produce velocity-temperature modes than can be compared directly to their energy-optimized counterparts, in terms of both the per-mode large scale organization and the convergence of energy, dissipation and heat flux globally and in the boundary layers.

The remainder of this paper is laid out as follows. In Section~\ref{sec:formalism} we present the generic formalism for POD and the particular formulations for the energy and dissipation optimized versions, as well as the data set upon which the subsequent analysis is based. The data set is symmetrized based on the geometrical symmetries of the cell, and we analyze the formal consequences that this has on the decomposition in Section~\ref{sec:symmetries}. In Section~\ref{sec:pod_specta_modes} we analyze the results of the energy and dissipation decomposition, including spectra, large-scale organization of POD modes, and resolved boundary layer structures. The convergence of reconstructed quantities in considered in Section~\ref{sec:convergence}. We discuss the implications of the results for the applicability of the formalism in a modeling context in Section~\ref{sec:modeling_perspectives}, and summarize the conclusions in Section~\ref{sec:conclusion}.

\section{POD Formalism\label{sec:formalism}}
POD was introduced to turbulence studies by \citet{lumley1967structure} as a method for educing large-scale coherent structures in turbulent flows. It has found use both as a valuable analytical tool for understanding mechanisms in flow dynamics (see \citet{bakewell1967viscous} and \citet{herzog1986large} for early examples), as a modeling component in the context of reduced-order models (ROMs) \citep{aubry1988dynamics,deane1991low}, and for flow control\citep{ravindran2000reduced,ly2001modeling}. In general terms, POD produces an orthogonal basis for a given data set such that the convergence of an expansion of the data set in this basis is optimal in the mean with respect to the norm on the data set. Formally, we let the data set $\mathcal{Q} = \{q_m\}_{m=1}^M$ of size $M$ be a subset of a Hilbert space $\mathcal{H}$ with the associated inner product $\ip{\cdot}{\cdot}{}$ and norm $\left\lVert\cdot\right\rVert=\sqrt{\ip{\cdot}{\cdot}{}}$. The data set is equipped with an averaging operation, $\mean{\cdot_m}{m=1}{M}$. Each mode in the resulting orthogonal modal basis $\{\varphi_n\}_{n=1}^N$ maximizes the mean projection of the data set,
\begin{align}
    \varphi_n &= \argmax_{\varphi\in\mathcal{H}_n} \frac{\mean{\left|\ip{\varphi}{q_m}{}\right|}{m=1}{M}}{\left\lVert\varphi\right\rVert^2}\,,\quad n=1, 2, \ldots, N\,,
    \label{eq:pot_optimization}
\end{align}
where $\mathcal{H}_n = \mathcal{H}\setminus\spn{\left(\{\varphi_{n'}\}_{n'=1}^{n-1}\right)}$, and $N$ is the number of modes needed to span the data set, i.e., its effective dimensionality, which satisfies $N\leq M$.

The POD modes are eigenmodes of the POD operator $R$,
\begin{align}
    R\varphi_n &= \lambda_n \varphi_n\,, \quad R\varphi = \mean{\ip{\varphi}{q_m}{}q_m}{m=1}{M}\,.
    \label{eq:pod_evp}
\end{align}
This operator is Hermitian and positive semi-definite by construction, ensuring that the eigenmodes resulting from \eqref{eq:pod_evp} form a complete orthogonal basis for the data set and that the eigenvalues are real and non-negative. The eigenpairs are conventionally ordered by decreasing eigenvalues, corresponding to modes being ordered by decreasing mean amplitude. The expansion of elements of the data set,
\begin{align}
    q_m &= \sum_{n=1}^N a_{nm} \sigma_n \varphi_n\,,
    \label{eq:pod_expansion}
\end{align}
where $\sigma_n = \sqrt{\lambda_n}$ are the singular values, and expansion coefficients are given by
\begin{align}
    a_{nm} &= \frac{1}{\sigma_n}\ip{\varphi_n}{q_m}{}\,,
    \label{eq:pod_coefficient}
\end{align}
is thus optimally robust against truncation, preserving as much of the mean norm of the truncated expansion as is possible for any basis. The eigenvalue $\lambda_n$ is the mean energy resolved by the mode $\varphi_n$, which can be interpreted as a measure of the importance of the mode in the reconstruction of the data set. It should be noted, however, that eigenvalues measure the importance of modes only in the narrow sense defined by the inner product; it is thus entirely possible for features of central importance for the dynamics of the flow to be resolved only by modes with low eigenvalues. One consequence of this is the difficulty of constraining \emph{a priori} the order of POD-based ROMs needed to achieve the desired model accuracy.

 The POD coefficients $a_{nm}$ in \eqref{eq:pod_expansion} are uncorrelated,
\begin{align}
    \mean{a_{nm}a_{n'm}}{m=1}{M} &= \delta_{nn'}\,,
    \label{eq:uncorrelated_coeffs}
\end{align}
where $\delta_{nn'}$ denotes the Kronecker delta. Different normalization conventions exist regarding the definition of POD coefficients and the expansion in \eqref{eq:pod_expansion}, with singular values often absorbed in coefficients. In that case $\sigma_n$ disappears from \eqref{eq:pod_expansion} and \eqref{eq:pod_coefficient}, and the right-hand side in \eqref{eq:uncorrelated_coeffs} becomes $\lambda_n \delta_{nn'}$.

The normalization convention chosen in the present work emphasizes the analogy between the expansion in \eqref{eq:pod_expansion} and the singular value decomposition (SVD). The scaled data matrix $Q\in\mathbb{R}^{N_f\times M}$ is formed with entries given by $Q_{ij} = \frac{1}{\sqrt{M}} q_j(x_i)$, where $N_f$ is the number of degrees of freedom in the flow field. The SVD of $Q$ consists of the decomposition
\begin{align}
    Q &= \Phi \Sigma A^{\mathsf{T}}\,,
    \label{eq:svd}
\end{align}
where $\Phi\in \mathbb{R}^{N_f\times N}$ and $A \in \mathbb{R}^{M\times N}$ are orthonormal matrices with POD modes and modal coefficients as their respective columns, and $\Sigma \in \mathbb{R}^{N\times N}$ is a diagonal matrix with the singular values along its diagonal. The eigenvalue problem in \eqref{eq:pod_evp} can then be compactly written as
\begin{align}
    QQ^{\mathsf{T}} \Phi &= \Phi \Lambda\,,
    \label{eq:svd_evp}
\end{align}
where $\Lambda = \Sigma\Sigma^{\mathsf{T}} \in \mathbb{R}^{N\times N}$ is a diagonal matrix with eigenvalues along its diagonal.

Commonly, $N_f \gg M$, and the $N_f\times N_f$ matrix $Q Q^{\mathsf{T}}$ becomes impractically large. The method of snapshots introduced by \citet{sirovich1987turbulence} replaces the matrix eigenvalue problem in \eqref{eq:svd_evp} with the snapshot eigenvalue problem,
\begin{align}
    Q^{\mathsf{T}} Q A &= A \Lambda\,,
    \label{eq:snapshot_matrix_evp}
\end{align}
where the matrix $Q^{\mathsf{T}} Q \in\mathbb{R}^{M\times M}$ presents a much more manageable eigenvalue problem. After solving \eqref{eq:snapshot_matrix_evp} for the coefficient matrix $A$ and eigenvalue matrix $\Lambda$ the mode matrix can be recovered as $\Phi = Q A \Sigma^{-1}$.

In the original formulation by \citet{lumley1967structure} the data set was an ensemble of independent flow realizations defined on a spatio-temporal domain. This method is in principle capable of capturing the full spatial and temporal dynamics governing the data set, although the size of the necessary data sets may make it prohibitively expensive for fully three-dimensional time-dependent flows. A number of more computationally tractable approaches have been suggested. Commonly, the ensemble is formed from a set of measurement or simulation snapshots, each representing a single instant in time. Only purely spatial correlations can be captured from such a data set, as all information relating to temporal structure is discarded in treating the time series as a statistical ensemble. \citet{zhang2023phase} demonstrated a domain combining spatial dimensions and phase for a two-dimensional periodic lid driven cavity flow, forming the ensemble from different periods of the flow. The combined spatial and temporal dynamics within a period could thus be described. \citet{lumley1967structure} showed how dimensions exhibiting homogeneity and periodicity could be eliminated from the domain by replacing $R$ with its Fourier transform along these dimensions, yielding a separate eigenvalue problem for each wave number or frequency. This approach has become known as spectral POD \citep{towne2018spectral} (not to be confused with the identically named method proposed by \citet{sieber2016spectral}). The method has also been used to approximate POD modes along dimensions that were not manifestly periodic, although this introduces certain deviations from the optimality offered by true POD, cf.~\citet{hodvzic2022discrepancies}. 

\subsection{Energy POD}
In this work we consider the decompositions based on two different ensembles: one formed from a set of snapshots of velocity and temperature data, leading to a decomposition optimized with respect to total energy (kinetic $+$ thermal); and one formed from snapshots of the corresponding strain rate tensor and temperature gradient data, optimized with respect to total (viscous $+$ thermal) dissipation. The velocity-temperature data set is described in more detail in Section~\ref{sec:data_set}. In either case the POD is fully specified by defining the ensemble $\mathcal{Q}=\{q_m\}_{m=1}^M$ and the inner product $\ip{\cdot}{\cdot}{}$ used in \eqref{eq:pot_optimization}, \eqref{eq:pod_evp}, and \eqref{eq:pod_coefficient}. 

For the energy POD the ensemble $\mathcal{Q}_E = \{q_m\}_{m=1}^M\subset\mathcal{H}_E$ consists of combined velocity and temperature snapshots,
\begin{align}
    q_m &= \left(u_m, v_m, w_m, \theta_m\right)\,,
    \label{eq:combined_vt_snapshot}
\end{align}
which are elements in the Hilbert space
\begin{align}
    \mathcal{H}_E &= \left\{\varphi: \Omega \rightarrow \mathbb{R}^4 \left| \ip{\varphi}{\varphi}{E} < \infty \right\}\right.\,.
\end{align}
This space is equipped with the inner product $\ip{\cdot}{\cdot}{E}$ and norm $\left\lVert\cdot\right\rVert_E$ corresponding to total (kinetic and thermal) energy, defined as
\begin{subequations}
    \begin{align}
        \ip{\varphi}{\psi}{E} &= \sum_{i=1}^3 \int_{\Omega} \varphi^i\psi^i \,\mathrm{d}x + \gamma_E^2 \int_{\Omega} \varphi^{\theta}\psi^{\theta}\,\mathrm{d}x\,, \label{eq:energy_ip}\\ \left\lVert\varphi\right\rVert_E&=\sqrt{\ip{\varphi}{\varphi}{E}}\,, \label{eq:energy_norm}
    \end{align}
    \label{eq:energy_ip_norm}
\end{subequations}
where $\varphi,\psi \in \mathcal{H}_E$. The first of the two terms in \eqref{eq:energy_ip} accounts for kinetic energy, and the second term accounts for thermal energy. The weighing factor $\gamma_{E}$ in \eqref{eq:energy_ip} is defined so as to ensure that kinetic and thermal energy contribute equally to the norm when averaged over the snapshot ensemble,
\begin{align}
    \gamma_E &= \left(\frac{\mean{\sum_{i=1}^3 \int_{\Omega} \left(q_m^i\right)^2 \,\mathrm{d}x}{m=1}{M}}{\mean{\int_{\Omega} \left(\theta_{m'} \right)^2\,\mathrm{d}x}{m'=1}{M}}\right)^{\frac{1}{2}}\,.
    \label{eq:gamma_E}
\end{align}

The total energy POD operator is then
\begin{align}
    R_E \varphi &= \mean{\ip{\varphi}{q_m}{E} q_m}{m=1}{M}\,.
    \label{eq:energy_pod_op}
\end{align}

Solutions to the eigenvalue problem resulting from setting $R=R_E$ in \eqref{eq:pod_evp} are temperature-velocity modes optimized with respect to the total energy. This was done by \citet{soucasse2019proper} for the data set also considered in this work, in which the results serve as a point of comparison for results of the dissipation POD described in the following.

\subsection{Dissipation POD}
The dissipation POD is based on the idea of letting mean dissipation take on the role played by mean energy in the energy POD formulated in the previous section. The mean dissipation in question is formed by combining mean viscous dissipation, $\left\langle\varepsilon_u\right\rangle$, and mean thermal dissipation, $\left\langle\varepsilon_{\theta}\right\rangle$. Our task is therefore to formulate the relevant ensemble, along with an inner product with an associated norm that produces the total dissipation when applied to the ensemble. For this purpose, we consider the strain rate tensor (SRT) and thermal gradient,
\begin{subequations}
    \begin{align}
        q_{m,\mathrm{srt}}^{ij} &=  \frac{1}{2} \left(\nabla^i q_m^j + \nabla^j q_m^i\right)\,,\\
        q_{m,\mathrm{tg}}^i &= \nabla^i \theta_{m}\,,
    \end{align}
    \label{eq:grad_ops}
\end{subequations}
with $i,j\in\{1,2,3\}$. From the SRT and thermal gradient we obtain the mean viscous and thermal dissipation, and, by combining these two, the total dissipation,
\begin{subequations}
    \begin{align}
        \left\langle\varepsilon_u \right\rangle &= \mean{\sum_{i,j=1}^3 \int_{\Omega}\left|q_{m,\mathrm{srt}}^{ij}\right|^2\,\mathrm{d}x}{m=1}{M}\,,\label{eq:viscous_dissipation}\\
        \left\langle\varepsilon_{\theta}\right\rangle &= \mean{\sum_{i=1}^3 \int_{\Omega} \left|q_{\mathrm{tg},m}^i\right|^2 \,\mathrm{d}x}{m=1}{M}\,, \label{eq:thermal_dissipation}\\
        \left\langle\varepsilon_{\mathrm{tot}}\right\rangle &= \left\langle\varepsilon_u\right\rangle + \gamma_D^2\left\langle\varepsilon_{\theta}\right\rangle\,,\label{eq:total_dissipation}
    \end{align}
    \label{eq:dissipation_first}
\end{subequations}
where, similarly to $\gamma_E$ in \eqref{eq:gamma_E}, $\gamma_{D}$ in \eqref{eq:total_dissipation} ensures equal mean contribution from viscous and thermal dissipation,
\begin{align}
    \gamma_{D} &= \left(\frac{\left\langle\varepsilon_u\right\rangle}{\left\langle\varepsilon_{\theta}\right\rangle}\right)^{\frac{1}{2}}\,.
    \label{eq:gamma_D}
\end{align}

The dissipation POD ensemble $\mathcal{Q}_D = \{q_m'\}_{m=1}^M\subset \mathcal{H}_D$ is formed by joining the objects defined in \eqref{eq:grad_ops},
\begin{align}
    q_m' &= \mathfrak{D} q_m = \left(q_{m,\mathrm{srt}}, q_{m,\mathrm{tg}}\right)\,,
    \label{eq:srttg_snapshot}
\end{align}
to form elements in the Hilbert space $\mathcal{H}_D = \mathcal{H}_{\mathrm{srt}}\times\mathcal{H}_{\mathrm{tg}}$. Here, $\mathcal{H}_{\mathrm{srt}}$ and $\mathcal{H}_{\mathrm{tg}}$ are Hilbert spaces containing $q_{m,\mathrm{srt}}$ and $q_{m,\mathrm{tg}}$, respectively,
\begin{subequations}
    \begin{align}
        \mathcal{H}_{\mathrm{srt}} &= \left.\left\{\varphi:\Omega\rightarrow \mathbb{R}^{3\times 3} \right| \ip{\varphi}{\varphi}{\mathrm{srt}} < \infty \right\}\,,\\
        \mathcal{H}_{\mathrm{tg}} &= \left.\left\{\varphi:\Omega\rightarrow \mathbb{R}^3 \right| \ip{\varphi}{\varphi}{\mathrm{tg}} < \infty \right\}\,,
    \end{align}
    \label{eq:grad_spaces}
\end{subequations}
where the inner products $\ip{\cdot}{\cdot}{\mathrm{srt}}$ and $\ip{\cdot}{\cdot}{\mathrm{tg}}$ and their associated norms are
\begin{subequations}
    \begin{align}
        \ip{\varphi_{\mathrm{srt}}}{\psi_{\mathrm{srt}}}{\mathrm{srt}} &= \sum_{i,j=1}^3\int_{\Omega} \varphi^{ij}_{\mathrm{srt}}\psi^{ij}_{\mathrm{srt}}\,\mathrm{d}x\,,\\
        \left\lVert\varphi_{\mathrm{srt}}\right\rVert_{\mathrm{srt}} &= \sqrt{\ip{\varphi_{\mathrm{srt}}}{\varphi_{\mathrm{srt}}}{\mathrm{srt}}}\,;\\
        \ip{\varphi_{\mathrm{tg}}}{\psi_{\mathrm{tg}}}{\mathrm{tg}} &= \sum_{i=1}^3\int_{\Omega}\varphi^i_{\mathrm{tg}}\psi^j_{\mathrm{tg}}\,\mathrm{d}x\,,\\
        \left\lVert\varphi_{\mathrm{tg}}\right\rVert_{\mathrm{tg}} &= \sqrt{\ip{\varphi_{\mathrm{tg}}}{\varphi_{\mathrm{tg}}}{\mathrm{tg}}}\,,
    \end{align}
\end{subequations}
where $\varphi_{\mathrm{srt}},\psi_{\mathrm{srt}}\in\mathcal{H}_{\mathrm{srt}}$ and $\varphi_{\mathrm{tg}},\psi_{\mathrm{tg}}\in\mathcal{H}_{\mathrm{tg}}$. The operator \mbox{$\mathfrak{D}: \mathcal{H}_E\rightarrow\mathcal{H}_D$} in \eqref{eq:srttg_snapshot} maps velocity-temperature snapshots to the corresponding joint SRT and thermal gradients.

The total dissipation inner product and norm can now be written as
\begin{subequations}
    \begin{align}
        \ip{\varphi'}{\psi'}{D} &= \ip{\varphi_{\mathrm{srt}}}{\psi_{\mathrm{srt}}}{\mathrm{srt}} + \gamma_D^2 \ip{\varphi_{\mathrm{tg}}}{\psi_{\mathrm{tg}}}{\mathrm{tg}}\,,\\
        \left\lVert \varphi'\right\rVert_D &= \sqrt{\ip{\varphi'}{\varphi'}{D}}\,,
    \end{align}
\end{subequations}
with $\varphi',\psi'\in\mathcal{H}_D$. The total dissipation POD operator is
\begin{align}
    R_{D} \varphi' &= \mean{\ip{\varphi'}{q_m'}{D} q_m'}{m=1}{M}\,.
    \label{eq:dissipation_pod_op}
\end{align}

The decomposition of $\mathcal{Q}_D$ which is obtained using this operator is optimized with respect to the norm $\left\lVert\cdot\right\rVert_D$ corresponding to total dissipation,
\begin{align}
    \left\langle\varepsilon_{\mathrm{tot}}\right\rangle &= \mean{\left\lVert q'_m\right\rVert_D}{m=1}{M}\,.
\end{align}

\subsection{The extended snapshot method}
The POD eigenvalue problems derived from the operators defined in \eqref{eq:energy_pod_op} and \eqref{eq:dissipation_pod_op} are
\begin{subequations}
    \begin{align}
        R_{E} \varphi_{E,n} &= \lambda_{E,n} \varphi_{E,n}\,, \label{eq:energy_space_evp}\\
        R_{D} \varphi_{D,n}' &= \lambda_{D,n}\varphi_{D,n}'\,. \label{eq:diss_space_evp}
    \end{align}
    \label{eq:space_evp}
\end{subequations}
The solutions of \eqref{eq:energy_space_evp} are energy eigenmodes which span $\mathcal{Q}_E$ optimally with respect to the inner product $\ip{\cdot}{\cdot}{E}$, whereas solutions of \eqref{eq:diss_space_evp} are dissipation eigenmodes spanning $\mathcal{Q}_D$ optimally with respect to $\ip{\cdot}{\cdot}{D}$. The eigenmodes in the energy basis $\{\varphi_{E,n}\}_{n=1}^N$ and those in the dissipation basis $\{\varphi_{D,n}'\}_{n=1}^N$ are elements of different Hilbert spaces, and cannot be compared directly.

We generate velocity-temperature modes $\{\varphi_{D,n}\}_{n=1}^N$ corresponding to $\{\varphi_{D,n}'\}_{n=1}^N$ using the extended POD method introduced by \citet{boree2003extended}, combined with the snapshot POD formalism of \citet{sirovich1987turbulence}. The resulting procedure is similar to what was done by \citet{podvin2015large}, as well as formally equivalent to the procedure presented by \citet{olesen2023dissipation}. In the snapshot formalism we construct the snapshot matrices corresponding to $Q^{\mathsf{T}}Q$,
\begin{subequations}
    \begin{align}
        \left[S_E\right]_{m,m'} &= \frac{1}{M} \ip{q_m}{q_{m'}}{E}\,, \\
        \left[S_{D}\right]_{m,m'} &= \frac{1}{M} \ip{q_m'}{q_{m'}'}{D}\,,
    \end{align}
    \label{eq:snapshot_matrices}
\end{subequations}
and solve the corresponding eigenvalue problems for the coefficient vectors $\{a_{p,n}\}_{n=1}^N$,
\begin{align}
    S_p a_{p,n} &= \lambda_{p,n} a_{p,n}\,,\quad p\in\{E,D\}\,.
    \label{eq:snapshot_evp}
\end{align}

Velocity-temperature modes are then constructed using the coefficients,
\begin{align}
    \varphi_{p,n} &= \frac{1}{\sigma_{p,n}} \mean{a_{p,n,m} q_m}{m=1}{M}\,,\quad p \in\{E,D\}\,,
    \label{eq:veltemp_modes}
\end{align}
where $\sigma_{p,n} = \sqrt{\lambda_{p,n}}$.

The basis $\{\varphi_{E,n}\}_{n=1}^N$ formed this way is identical to the eigenbasis obtained from \eqref{eq:energy_space_evp}. In contrast, $\{\varphi_{D,n}\}_{n=1}^N$ is \emph{not} the eigenbasis obtained from \eqref{eq:diss_space_evp}; the modes satisfy $\varphi_{D,n}' = \mathfrak{D}\varphi_{D,n}$, and form a complete (but non-orthogonal) basis for the data set. The data set can therefore be reconstructed using either basis as, cf.~\eqref{eq:pod_expansion},
\begin{align}
    q_m &= \sum_{n=1}^N a_{p,n,m} \sigma_{p,n} \varphi_{p,n}\,,\quad p\in\{E,D\}\,.
    \label{eq:data_reconstruction}
\end{align}

Both coefficient sets are uncorrelated, cf.~\eqref{eq:uncorrelated_coeffs},
\begin{align}
    \mean{a_{p,n,m} a_{p,n',m}}{m=1}{M} &= \delta_{nn'}\,,\quad p\in\{E,D\}\,,
    \label{eq:uncorrelated_coeff}
\end{align}
allowing reconstruction of any second order mean quantity (such as energy, dissipation, heat flux, and stresses) without the use of cross-modal terms. This lets us decompose mean quantities into a sum of modal contributions,
\begin{align}
    \Pi^p_n &= \frac{1}{M}\sum_{n'=1}^n \lambda_{p,n'}\Pi\left[\varphi_{p,n'}\right]\,,
    \label{eq:profile_convergence}
\end{align}
where $\Pi\in\{E_u, E_{\theta}, \varepsilon_u, \varepsilon_{\theta}, \Phi\}$ identifies the reconstructed quantity (kinetic and thermal energy, viscous and thermal dissipation, and convective heat flux), $p\in\{E,D\}$ designates the decomposition, $n$ the order of the reconstruction (number of modes included), and $\Pi[\varphi]$ denotes the reconstructed quantity computed from mode $\varphi$. Mean quantities are recovered as $\left\langle \Pi\right\rangle = \Pi^p_N$, which due to the completeness of either basis is independent of $p$. This makes it possible to identify the contribution from each mode to each mean profile, and, as we shall see, to link this to the structures described by the mode.

\subsection{Velocity and temperature fields\label{sec:data_set}}
The numerical setup and associated data set are the same as those used in~\citet{soucasse2019proper}. The configuration studied is a cubic Rayleigh-Bénard cell filled with air, with isothermal horizontal walls and adiabatic side walls. The air is assumed to be transparent and thermal radiation effects are disregarded. Direct numerical simulations have been performed at a Rayleigh number $\mathrm{Ra}=10^7$. The Prandtl number $\mathrm{Pr}=\nu/\kappa$ is set to 0.707.  All physical quantities are made dimensionless using the cell size $H$, the reference time $H^2/(\kappa\sqrt{\mathrm{Ra}})$ and the reduced temperature $\theta=(T-T_0)/\Delta T$, $T_0$ being the mean temperature between hot and cold walls. Spatial coordinates are denoted $x$, $y$, $z$ ($z$ being the vertical direction) and the origin is placed at a bottom corner of the cube.

The Navier–Stokes equations under the Boussinesq approximation are solved using a Chebyshev collocation method~\citep{xin02,xin-PCFD08}. Computations are made parallel using domain decomposition along the vertical direction. Time integration is performed through a second-order semi-implicit scheme. The velocity divergence-free condition is enforced using a projection method. The spatial mesh is made of 81 Chebyshev collocation points in each direction of space. We have checked that the number of collocation points is sufficient to accurately discretize the boundary layers according to the criterion proposed by~\citet{shishkina2010boundary}. The resulting Nusselt number averaged over time and over the horizontal plane is $\mathrm{Nu}=16.24$, and the resulting Reynolds number based on the volume and time average of the velocity magnitude is
\begin{align}
    \mathrm{Re} &= \frac{\sqrt{\left\langle u_i u_i\right\rangle_{v,t} \mathrm{Ra}}}{\mathrm{Pr}} = 651\,.
    \label{eq:reynolds}
\end{align}
These values are in good agreement with reference literature results~\citep{delort2022rayleigh}.

\begin{figure*}
	\includegraphics{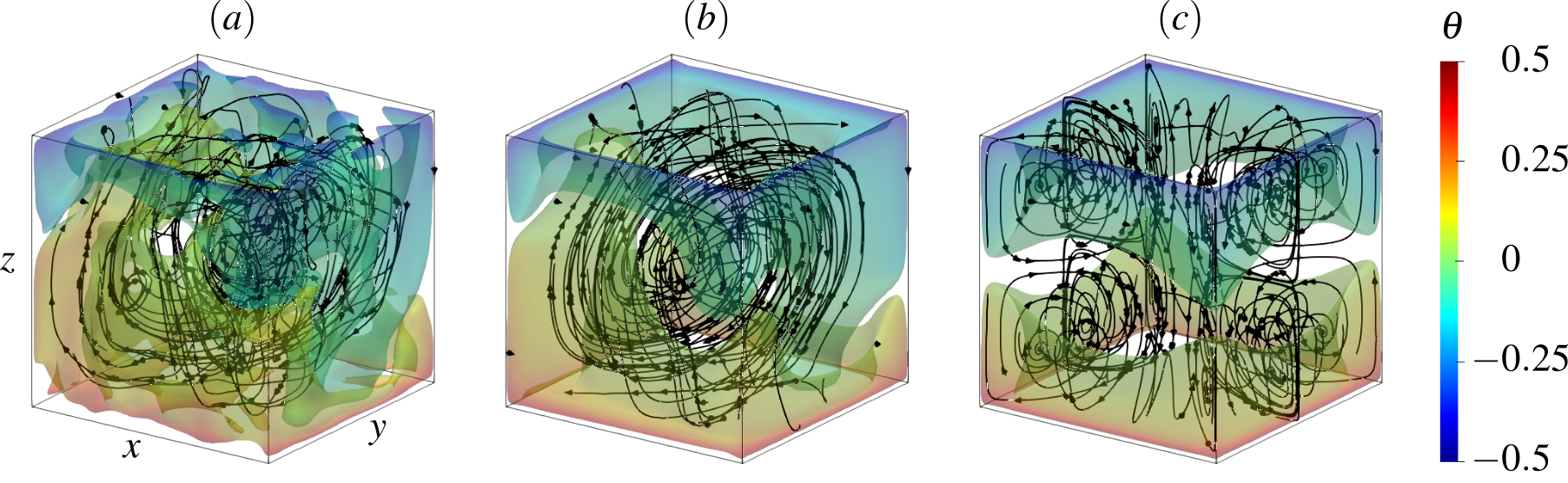}
    \caption{The snapshot $m=1$ $(a)$, the conditional mean field given $L_x < 0$, $L_y > 0$ $(b)$, and the full mean field $(c)$, shown as isosurfaces of scaled convective heat flux corresponding to $\Phi_{\mathrm{iso}} = 0.05\cdot \max_{\Omega}\left(w\theta\right)$ and colored by temperature $\theta$. In this and the similar plots in Figures~\ref{fig:similar_modes_1} and~\ref{fig:similar_modes_2} the origin is located in the lower left corner of the cell, and the axes are labeled as shown in $(a)$.
    \label{fig:mean_snap}}
\end{figure*}

A number of 1000 snapshots has been extracted from the simulation at a sampling period of 10 in dimensionless time units. Isosurfaces of convective heat flux and stream lines are shown in Figure~\ref{fig:mean_snap} for the first snapshot (Figure~\ref{fig:mean_snap}$a$), for the conditional mean of realizations with a particular orientation of integrated angular momentum ($L_x < 0$, $L_y > 0$; Figure~\ref{fig:mean_snap}$b$), and for the full mean (Figure~\ref{fig:mean_snap}$c$) of the symmetry-enriched data set, cf.~Section~\ref{sec:symmetries}. The isosurfaces highlight cold ($\theta < 0$) and warm ($\theta > 0$) convective regions, where appreciable net upward convective heat transfer ($w\theta > 0.05$) occur due to warm fluid moving upwards in the case of warm convective regions, and cold fluid moving downwards in the case of cold convective regions. Figure~\ref{fig:mean_snap}$a,b$ shows a diagonally aligned roll producing oppositely oriented convective regions on each side. This represents the typical overall organization of the flow, with the orientation of the roll shifting with irregular intervals. This contrasts with the mean flow shown in Figure~\ref{fig:mean_snap}$c$, which does not represent a typical flow state.

\section{Symmetries\label{sec:symmetries}}
As in~\citet{soucasse2019proper} the data set is enriched by making use of the statistical symmetries of the flow~\citep{puigjaner2008bifurcation}. In the cubic Rayleigh-Bénard cell, four quasi-stable states are available for the flow at this Rayleigh number: the large-scale circulation (LSC) settles in one of the two diagonal planes of the cube with clockwise or counterclockwise motion. Reorientations from one state to another occur during the analyzed time sequence, but each state is not necessarily equally visited. In order to counteract this bias we construct an enlarged snapshot ensemble, obtained by the action of the symmetry group of the problem on the original ensemble.

The symmetry group is formed from the four basic symmetry operations $S_{k\in\{x,y,z,d\}}$\citep{puigjaner2008bifurcation},
\begin{subequations}
    \begin{align}
        S_x&:\left\{\begin{array}{rcl}
         \left(x,y,z\right)&\mapsto& \left(1-x, y, z\right)  \\
          \left(u,v,w,\theta\right) &\mapsto& \left(-u,v,w,\theta\right)
        \end{array}
        \right.\,,\label{eq:sym_x}\\
        S_y&:\left\{\begin{array}{rcl}
         \left(x,y,z\right)&\mapsto& \left(x, 1-y, z\right)  \\
          \left(u,v,w,\theta\right) &\mapsto& \left(u,-v,w,\theta\right)
        \end{array}
        \right.\,,\label{eq:sym_y}\\
        S_z&:\left\{\begin{array}{rcl}
         \left(x,y,z\right) &\mapsto& \left(x, y, 1-z\right)  \\
          \left(u,v,w,\theta\right) &\mapsto& \left(u,v,-w,-\theta\right)
        \end{array}
        \right.\,,\label{eq:sym_z}\\
        S_d&:\left\{\begin{array}{rcl}
         \left(x,y,z\right) &\mapsto& \left(y, x, z\right)  \\
          \left(u,v,w,\theta\right) &\mapsto& \left(v,u,w,\theta\right)
        \end{array}
        \right.\,.\label{eq:sym_d}
    \end{align}
    \label{eq:symmetries}
\end{subequations}

The operations $S_x$, $S_y$, and $S_z$ describe mirroring in the planes $x=0.5$, $y=0.5$, and $z=0.5$, respectively. This involves flipping the sign of the corresponding velocity component, and, for $S_z$, also the sign of the temperature fluctuation, due to the opposite thermal boundary conditions imposed at the top and bottom by the cold and the hot plates. The operation $S_d$ describes reflection in the diagonal plane $x=y$. The four symmetry operations form a 16-element symmetry group $\mathcal{G}$, and the enriched data set of $M=\num{16000}$ realizations is formed from the images of each of the $\widetilde{M} = \num{1000}$ DNS realizations $\{\tilde{q}_m\}_{m=1}^{\widetilde{M}}$ under each of the 16 symmetry group elements,
\begin{align}
    \left\{q_m\right\}_{m=1}^{M} &= \bigcup_{S\in\mathcal{G}}\left\{S \tilde{q}_{m'}\right\}_{m'=1}^{\widetilde{M}}\,.
\end{align}

The symmetries in the data set lead to degeneracies in the resulting POD bases. Here we show the formal origin and nature of these degeneracies using the commutation relations involving the POD and symmetry operators. Since the enriched data set is by construction invariant under each element in the symmetry group, the POD operators $R_{p\in\{E,D\}}$ commute with the symmetry operations. This implies that each POD operator shares an orthonormal eigenbasis with each symmetry operation, such that for any symmetry $S_k$ a POD eigenbasis $\{\varphi_{p,n}\}_{n=1}^N$ can be chosen satisfying $S_k \varphi_{p,n} = \mu_{k,p,n} \varphi_{p,n}$. The symmetry operations are unitary such that $\mu_{k,p,n} = \pm 1$, with the eigenvalue $+1$ indicating symmetry and $-1$ indicating antisymmetry with respect to $S_k$. By combining \eqref{eq:sym_x} and \eqref{eq:sym_d} we find
\begin{subequations}
    \begin{align}
        S_x S_d &: \left\{\begin{array}{rcl}
             \left(x,y,z\right) &\mapsto& \left(1-y, x, z\right)  \\
             \left(u,v,w,\theta\right) &\mapsto& \left(-v,u,w,\theta\right)
        \end{array}
        \right.\,,\\
        S_d S_x &: \left\{\begin{array}{rcl}
             \left(x,y,z\right) &\mapsto& \left(y, 1-x, z\right)  \\
             \left(u,v,w,\theta\right) &\mapsto& \left(v,-u,w,\theta\right)
        \end{array}
        \right.\,,
    \end{align}
\end{subequations}
implying that $\left[S_x,S_d\right]\neq 0$, and by a similar argument, $\left[S_y, S_d\right]\neq0$. From this it follows that $S_x$ and $S_y$ do not share orthogonal eigenbases with $S_d$, which in turn implies that the POD eigenbases are not unique, and that some form of degeneracy must exist.

We can show that $S_x S_d = S_d S_y$ and $S_y S_d = S_d S_x$, from which it follows that $[S_x + S_y, S_d] = 0$. Simultaneous eigenmodes of $S_x$ and $S_y$ satisfying $\mu_{x,p,n} = \mu_{y,p,n}$ are therefore also eigenmodes of $S_d$. Taken together, this means that degenerate POD eigenspaces of dimension two exist and are spanned by modes with $\mu_{x,p,n} = -\mu_{y,p,n}$. These degenerate pairs appear as pairs of repeated eigenvalues in the POD spectra, as shown in Figure~\ref{fig:spectra} in Section~\ref{sec:pod_specta_modes}.

The pair of degenerate POD eigenmodes spanning a given eigenspace can be chosen arbitrarily within that eigenspace, under the constraint of orthonormality. Each degenerate pair can thus be formed from one mode symmetric under $S_x$ ($\mu_{x,p,n}=+1$) and antisymmetric under $S_y$ ($\mu_{y,p,n} = -1$), and one with the opposite configuration, $\mu_{x,p,n'}=-1$ and $\mu_{y,p,n'} = +1$. These configurations correspond to modes which are oriented along the $x$ and $y$ axis, respectively, and which are not eigenmodes of $S_d$. The degeneracy is responsible for resolving the different roll orientations, and modes in degenerate eigenspaces are capable of supporting non-vanishing  angular momenta in the $xy$ plane, though it is not necessary that they do so. In this work we use the eigenmodes of $S_x$ and $S_y$ to span degenerate eigenspaces, resulting in modal angular momenta aligned with the $x$ and $y$ axes. One could equally well have spanned the eigenspaces using pairs of $S_d$ eigenmodes with $\mu_{d,p,n} = + 1$ and $\mu_{d,p,n'} = -1$, aligning the modal angular momenta with the diagonals. These modes would not be eigenmodes of $S_x$ and $S_y$. While this choice would arguably have been in closer agreement with the physical orientations attained by the flow (cf.~Figure \ref{fig:mean_snap}$a,b$, showing a typical diagonal-aligned state of the flow), the choice of aligning degenerate pairs along the principal axes allows for direct comparison with modes identified by \citet{soucasse2021low}. Since $\mu_{d,p,n}$ is undefined whenever $\mu_{x,p,n}\neq\mu_{y,p,n}$ we are left with 12 possible combinations of symmetry eigenvalues, or 12 distinct isosymmetric families. Within the lowest 20 POD modes we find each of those families represented by at least one mode, and in some cases by several modes.

The symmetry operations commute with the mapping $\mathfrak{D}$ from velocity-temperature snapshots to SRT-thermal gradient snapshots in \eqref{eq:srttg_snapshot}, and the preceding discussion therefore applies to both the POD eigenmodes and to the dissipation modes formed using \eqref{eq:veltemp_modes}. 

\section{POD spectra and modes\label{sec:pod_specta_modes}}
\begin{figure}
	
	\includegraphics{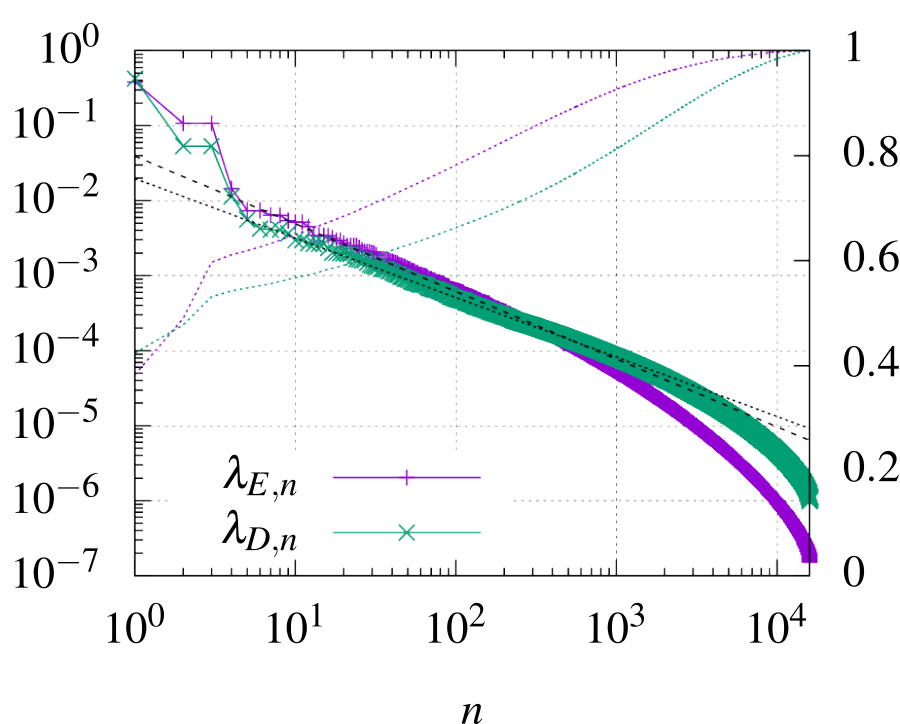}
    \caption{POD spectra (solid lines, left axis) and cumulative spectra (dotted lines, right axis) for PODs optimized with respect to total kinetic and thermal energy ($\lambda_{E,n}$) and total viscous and thermal dissipation ($\lambda_{D,n}$). The spectra are normalized such that $\sum_n \lambda_{E/D, n} = 1$. Power laws fitted to the spectra in the interval $10\leq n\leq 10^3$ are shown for the energy (black dashed line) and dissipation spectra (black dotted line).}
    \label{fig:spectra}
\end{figure}

Figure \ref{fig:spectra} shows the POD spectra corresponding to the two decompositions. It should be emphasized that while the energy eigenvalue $\lambda_{E,n}$ is the mean \emph{energy} contribution of the energy mode $\varphi_{E,n}$, the dissipation eigenvalue $\lambda_{D,n}$ is the mean \emph{dissipation} contribution of the dissipation mode $\varphi_{D,n}$. Consequently, as the eigenvalues of the different decompositions correspond to different norms, they are not directly comparable. Both spectra are dominated by the lowest three modes, representing circa 60\% of the total energy for the energy decomposition, and 55\% of the total dissipation for the dissipation decomposition. The remaining parts of the spectra decay slowly, with the energy POD spectrum exhibiting a slightly faster rate of decay than the dissipation POD spectrum. With $n=100$ modes we thus capture about 80\% of the total energy and 65\% of the total dissipation. The asymptotic decay rates are determined by fitting the spectra to power laws, $\lambda_{p,n}\sim n^{\alpha_p}$, in the interval $10\leq n \leq 10^3$. This yields the exponents $\alpha_E = -0.91$ and $\alpha_D = -0.79$ for the energy and dissipation spectra, respectively, confirming the more rapid convergence of the energy POD spectrum.

\subsection{Identification of modes\label{sec:mode_labeling}}
In order to facilitate a subsequent comparison between the modes produced by the energy and dissipation decomposition we assign labels to the lowest 20 modes of each decomposition, adopting and extending the modal nomenclature introduced for the first 12 energy optimized modes by \citet{soucasse2019proper,soucasse2021low}. Each of the energy modes considered here is assigned a label $G[\varphi_{E,n_E}]$, based on similarities to modes considered by \citet{soucasse2019proper,soucasse2021low}, and the same label is then assigned to the dissipation mode $\varphi_{D,n_D}$ selected among the lowest 20 modes that maximizes the overlap defined using the energy inner product and norm in \eqref{eq:energy_ip_norm}:
\begin{align}
    \Gamma(n_E, n_D) &= \frac{\left|\ip{\varphi_{E,n_E}}{\varphi_{D,n_D}}{E}\right|}{\left\lVert\varphi_{D,n_D}\right\rVert_E}\,.
    \label{eq:overlap_def}
\end{align}

The overlap $\Gamma(n_E,n_D) \in [0;1]$ measures the degree of similarity between modes of the different decompositions, with $\Gamma(n_E,n_D) = 1$ in the case when modes are identical up to normalization of $\varphi_{D,n_D}$, while the overlap vanishes in case of orthogonal modes.

Table \ref{tab:modes_characteristics} summarizes all the modes investigated in this way in the present work, their assigned label, their angular momentum and symmetry properties, and the overlap of modes with $G[\varphi_{E,n_E}]=G[\varphi_{D,n_D}]$. The assignment of shared labels between decompositions is found to be largely consistent with shared large scale organization, as discussed in Section~\ref{sec:diss_modes}.

\renewcommand{\arraystretch}{1.2}
\begin{table}[ht]
    \centering
    \begin{tabular}{rc|cc|c|cccc|c}
        & I & \multicolumn{2}{c|}{II} & III & \multicolumn{4}{c|}{IV} & V \\
        & $G[\varphi]$ & $n_E$ & $n_D$ & $L_i\neq 0$ & $\mu_x$ & $\mu_y$ & $\mu_z$ & $\mu_d$ & $\Gamma(n_E,n_D)$ \\ 
        \hline
         & $M$  & \multicolumn{2}{c|}{1} & & $+$ & $+$ & $+$ & $+$ & 1.000 \\
         \multirow{2}{*}{$\biggl\{$}
         & $L_x$ & \multicolumn{2}{c|}{2} & $x$ & $+$ & $-$ & $-$ & & \multirow{2}{*}{0.9996} \\
         & $L_y$ & \multicolumn{2}{c|}{3} & $y$ & $-$ & $+$ & $-$ & & \\
         & $D$ & \multicolumn{2}{c|}{4} && $-$ &  $-$ & $+$ & $+$ & 0.9981 \\ \hline
         \multirow{2}{*}{$\biggl\{$}
         & $BL_x$ & 5 & 7 & $x$ & $+$ & $-$ & $-$ & & \multirow{2}{*}{0.9103}\\
         & $BL_y$ & 6 & 8 & $y$ & $-$ & $+$ & $-$ & &\\
         & $C$ & 7 & 5 && $+$ & $+$ & $+$ & $-$ & 0.9870\\
         & $C^{\ast}$ & 8 & 6 && $+$ & $+ $& $-$ & $-$ & 0.9642 \\\hline
         & $D^{\ast}$ & \multicolumn{2}{c|}{9} && $-$ & $-$ & $-$ & $+$ & 0.9129\\
         \multirow{2}{*}{$\biggl\{$}
         & $BL_x^{\ast}$ & \multicolumn{2}{c|}{10} && $+$ & $-$ & $+$ & & \multirow{2}{*}{0.9741}\\
         & $BL_y^{\ast}$ & \multicolumn{2}{c|}{11} && $-$ & $+$ & $+$ & & \\
         & $M^{\ast}$ & \multicolumn{2}{c|}{12} && $+$ & $+$ & $-$ & $+$ & 0.9297\\ \hline
         \multirow{2}{*}{$\biggl\{$}
         & $L_x^{\dagger}$ & 13 & 19 & $x$ & $+$ & $-$ & $-$ & & \multirow{2}{*}{\textbf{0.6037}}\\
         & $L_y^{\dagger}$ & 14 & 20 & $y$ & $-$ & $+$ & $-$ & & \\
         & $M^{\dagger}$ & 15 & 16 && $+$ & $+$ & $+$ & $+$ & 0.8949\\
         & $L_z$ & 16 & 17 & $z$ & $-$ & $-$ & $+$ & $-$ & \textbf{0.6461} \\\hline
         & $C^{\dagger}$ & 17 & 13 & & $+$ & $+$ & $+$ & $-$ & 0.9446 \\
         \multirow{2}{*}{$\biggl\{$}
         & $K_x^{\dagger}$ & 18 & 14 && $+$ & $-$ & $+$ & & \multirow{2}{*}{0.9092} \\
         & $K_y^{\dagger}$ & 19 & 15 && $-$ & $+$ & $+$ & & \\
         & $D^{\dagger}$ & 20 & 18 && $-$ & $-$ & $-$ & $-$ & 0.8413 \\\hline
    \end{tabular}
    \caption{Characteristics of the lowest 20 POD modes of either decomposition, $\varphi_{E,n_E}$ and $\varphi_{D,n_D}$ with $n_{E/D}\leq 20$. Brackets on the left designate degenerate pairs. Mode labels in group I are consistent with those previously introduced by \citet{soucasse2019proper,soucasse2020reduced,soucasse2021low}, with additional modes designated using "$\dagger$". Group II shows mode indices in the energy ($n_E$) and dissipation ($n_D$) optimized decompositions; a single number is given when $n_E=n_D$. Group III lists non-vanishing angular momentum components. Group IV shows symmetries ($+$) and anti-symmetries ($-$) with respect to each of the four basic symmetry operations $S_x$, $S_y$, $S_z$, and $S_d$, corresponding to eigenvalues $\mu_k$ of the symmetry operation, cf.~$S_k \varphi = \mu_k \varphi$. Group V shows the overlap of modes from the two decompositions as defined in \eqref{eq:overlap_def} and shown in Figure~\ref{fig:mode_overlaps}. Since the overlap is the same for each pair of modes in a degenerate set only a single value is given for those. Boldface text denotes outlying overlap values. Rows are grouped to aid reading.}
    \label{tab:modes_characteristics}
\end{table}

\begin{figure}
	\includegraphics{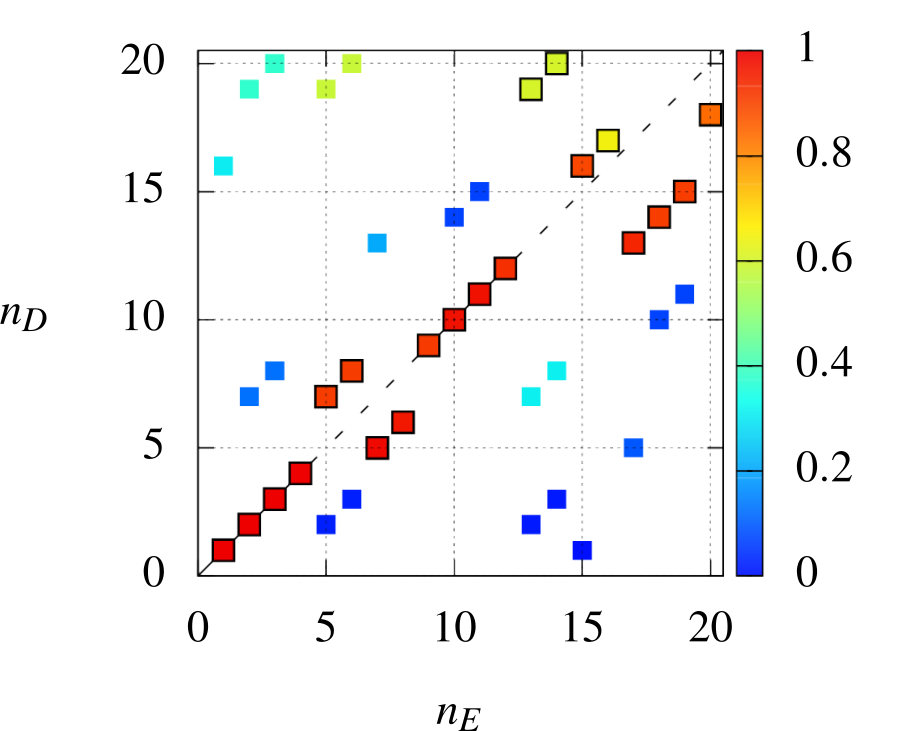}
    \caption{Non-vanishing overlap magnitudes $\Gamma(n_E, n_D)$ between energy and dissipation modes with $n_E,n_D \leq 20$, as defined in \eqref{eq:overlap_def}. Pairs with matching labels $G[\varphi_{E,n_E}] = G[\varphi_{D,n_D}]$ are marked with black frames, and the diagonal $n_E=n_D$ is shown as a dashed line. \label{fig:mode_overlaps}}
\end{figure}

The overlaps $\Gamma(n_E, n_D)$ for $n_E,n_D \leq 20$ are shown in Figure~\ref{fig:mode_overlaps}, where we have also highlighted pairs of modes for which $G[\varphi_{E,n_E}]=G[\varphi_{D,n_D}]$. Of such pairs, several modes ($n_{E/D}\leq 4$ and $9\leq n_{E/D} \leq 12$) preserve their indices between the decompositions, and exhibit large overlaps; these appear in Figure~\ref{fig:mode_overlaps} as framed red squares positioned on the diagonal. A number of modes are indexed differently, but still exhibit a large overlap (framed red squares off the diagonal); the significance of the reordering of energy modes 5--8 is discussed below. Lastly, a few of the modes considered show only moderate overlap (framed non-red squares). Non-framed squares denote overlaps between modes designated with different labels. The overlap matrix is sparse, as the overlap vanishes for mode pairs with non-identical symmetry signatures. Taken together, this points to a large degree of similarity between the decompositions, which will be investigated further in Section~\ref{sec:diss_modes}. It should be emphasized, however, that the similarity between the decompositions tends to diminish when considering modes higher than the ones investigated here.

A notable reordering of modes is the promotion of the $C$ and $C^{\ast}$ modes from indices 7 and 8 in the energy decomposition to indices 5 and 6 in the dissipation decomposition, ahead of the degenerate boundary roll mode pair $BL_{x/y}$. The promotion is moderately significant, in that the eigenvalue of the $BL_{x/y}$ modes exceed that of the $C$ modes by about 16\% in the energy decomposition, whereas the eigenvalue of the $C$ mode exceeds that of the $BL_{x/y}$ mode by $BL_{x/y}$ modes by 37\% in the dissipation decomposition. The $C$ mode was found by \citet{soucasse2019proper} to have a destabilizing effect on the LSC, and the promotion of this mode could thus be taken as an indication that certain dynamically important structures are captured more efficiently by the dissipation decomposition than by the energy decomposition.

For the modes $n_{E/D}\leq 20$ the ordering is to a large extent preserved between the decompositions, although this trend breaks down when considering higher modes. \citet{lee2020improving} found that ordering energy-optimized modes for a two-dimensional lid driven cavity flow by their modal contribution to mean enstrophy led to only a small amount of reordering compared to the native ordering of modes by energy, with the least amount of reordering found for low-index modes. Reordering was found to increase when increasing $\mathrm{Re}$. To the same effect, \citet{olesen2023dissipation} showed for a channel flow that both the mean dissipation contribution of energy modes and the mean energy contribution of dissipation modes tended to decrease with mode number. In both cases this was taken as an indication that the flow in question was as a whole dominated by a few structures spanning all relevant scales, and that the lowest POD modes were capturing structures on this range of scales. Such a lack of separation of scales could be a consequence of the moderate Rayleigh number of the flow investigated in this study.

\subsection{Large-scale organization of energy modes\label{sec:large_scale_organization}}
The energy decomposition of the data set used in this work has been considered by \citet{soucasse2019proper,soucasse2020reduced,soucasse2021low}, who described the structures contained in the lowest several modes along with the dynamics arising from a ROMs based on these modes. In this section we summarize and expand on the structural analysis, so as to gain a point of comparison for subsequent investigation of the corresponding dissipation modes.

\begin{figure*}
	\includegraphics{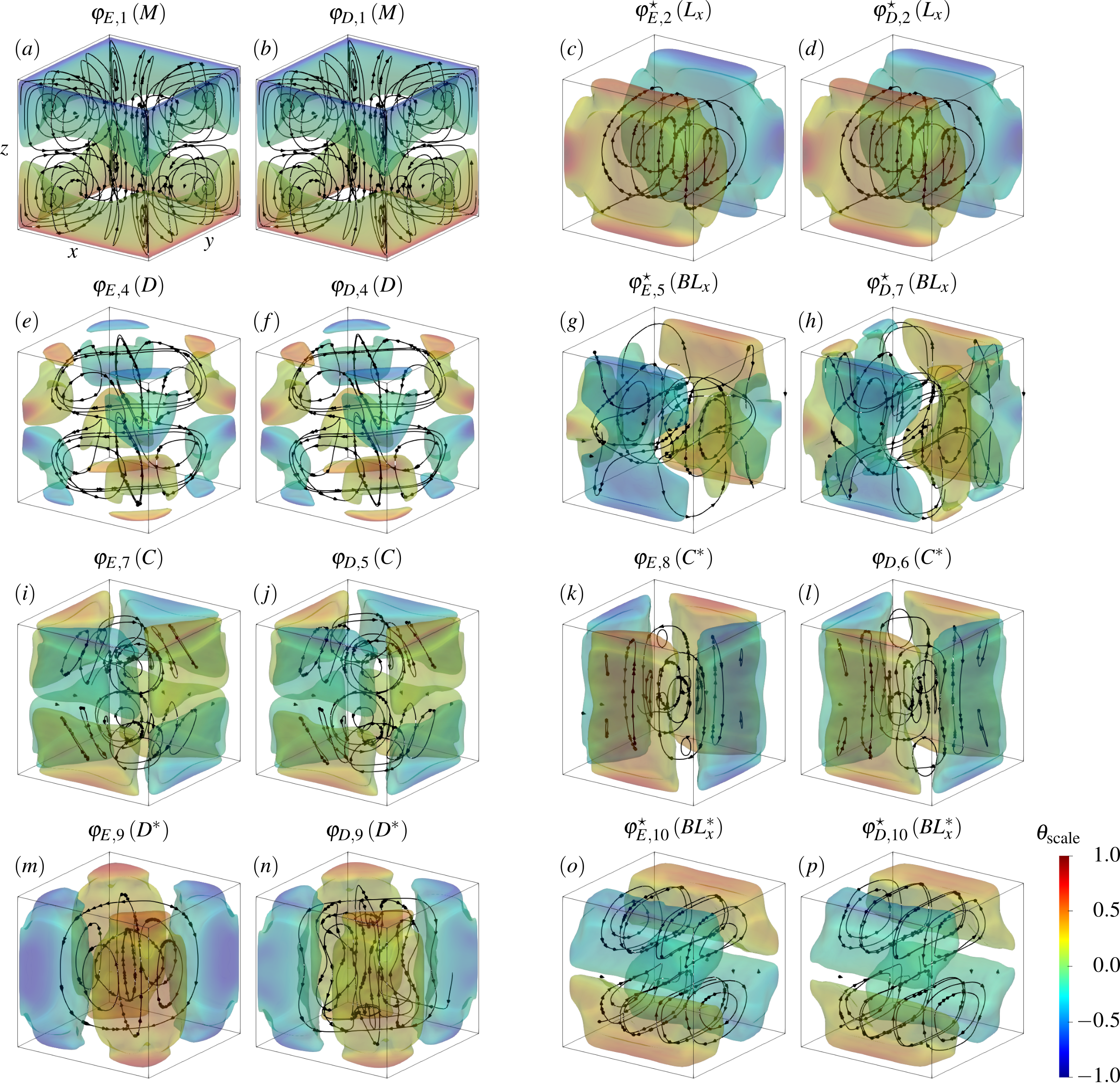}
   
    \caption{Energy-optimized POD modes $\varphi_{E,1}$--$\varphi_{E,10}$ juxtaposed with their similar dissipation-optimized POD modes $\varphi_{D,1}$--$\varphi_{D,10}$, shown as isocontours of scaled convective heat flux corresponding to $\Phi_{\mathrm{iso}} = 0.05\times \max_{\Omega}\left(\varphi^w \varphi^{\theta}\right)$ and colored by scaled temperature \mbox{$\theta_{\mathrm{scale}} = \varphi^{\theta}/\max_{\Omega}\left(\left|\varphi^{\theta}\right|\right)$}. For modes marked with $\star$ ($c,d,g,h,o,p$) only one mode of a degenerate pair is shown; the other mode in the pair is obtained by rotating the mode shown by $\frac{\pi}{2}$ around the central vertical axis.\label{fig:similar_modes_1}}
\end{figure*}

\begin{figure*}
	\includegraphics{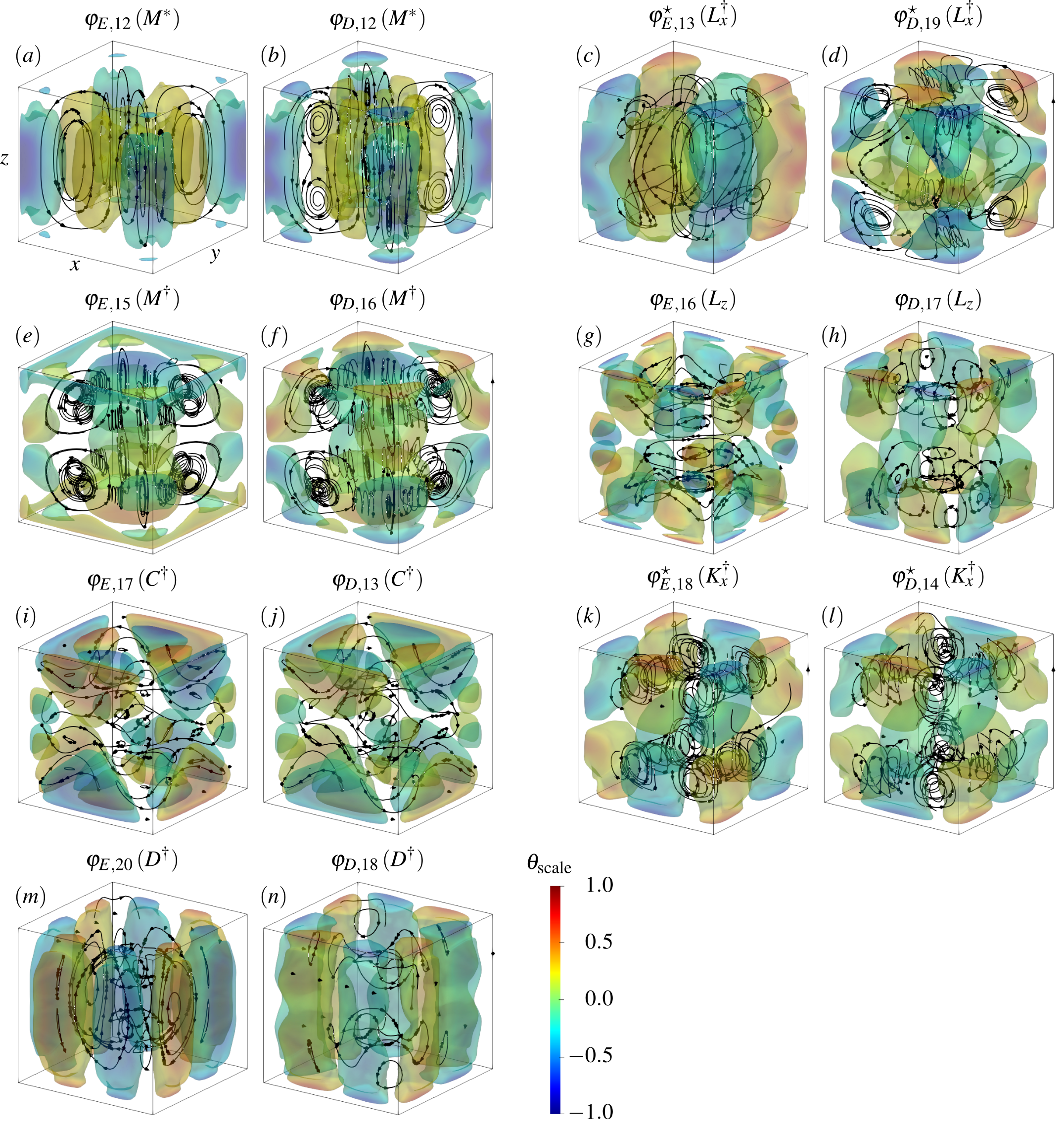}
    \caption{Energy-optimized POD modes $\varphi_{E,12}$--$\varphi_{E,20}$ juxtaposed with their similar dissipation-optimized POD modes $\varphi_{D,12}$--$\varphi_{D,20}$, shown as isocontours of scaled convective heat flux corresponding to $\Phi_{\mathrm{iso}} = 0.05\times \max_{\Omega}\left(\varphi^w \varphi^{\theta}\right)$ and colored by scaled temperature \mbox{$\theta_{\mathrm{scale}} = \varphi^{\theta}/\max_{\Omega}\left(\left|\varphi^{\theta}\right|\right)$}. For modes marked with $\star$ ($c,d,k,l$) only one mode of a degenerate pair is shown; the other mode in the pair is obtained by rotating the mode shown by $\frac{\pi}{2}$ around the central vertical axis. \label{fig:similar_modes_2}}
\end{figure*}

The large-scale organization of the lowest 20 modes of each decomposition is shown in Figures~\ref{fig:similar_modes_1}--\ref{fig:similar_modes_2} as temperature-colored heat flux isosurfaces along with stream lines. The lowest 20 energy modes are shown in order of increasing mode number, each juxtaposed with the dissipation mode sharing the label of the energy mode. Modes marked by "$^{\star}$" (not to be confused with "$^{\ast}$") represent one of a degenerate pair, for which the complimentary mode (not shown) can be obtained by rotating the mode shown by $\frac{\pi}{2}$ around the central vertical axis.

The $M$, $L_{x/y}$, $D$, $BL_{x/y}$ and $C$ modes ($\varphi_{E,1}$--$\varphi_{E,7}$) shown in Figure~\ref{fig:similar_modes_1}$a,c,e,g,i$, are identical to those obtained by \citet{soucasse2019proper}, who used the same data set as in the present work. The $M$ mode is approximately equal to the mean of the field taken over the full data set, covering all possible orientations of the LSC. It consists of a pair of counter-rotating toroidal structures which captures most of the thermal stratification. The $L_{x/y}$ modes form a degenerate pair, each mode describing a single roll along a principal axis; they combine to form the diagonal rolls characterizing the LSC. The $D$ mode consists of eight rolls, and transports fluid between the corners while stabilizing circulation along the diagonal. Together, the $M$, $L_{x/y}$, and $D$ modes form the LSC. The $BL_{x/y}$ and $C$ modes represent the dominant fluctuations superposed on the LSC. Each of the $BL_{x/y}$ modes corresponds to  a co-rotating pair of longitudinal rolls that connect the core to the boundary layers and modulate the $L_{x/y}$ modes. The $C$ mode corresponds to a corner structure that brings in fluid from one roll to another, and plays an important role in the LSC reorientations. The reader is referred to the paper by \citet{soucasse2019proper} for a more detailed discussion of these modes.

Additionally, modes similar to the $C^{\ast}$, $D^{\ast}$, $BL_{x/y}^{\ast}$, $M^{\ast}$, and $L_z$ modes ($\varphi_{E,8}$--$\varphi_{E,12}$ and $\varphi_{E,16}$), shown in Figures~\ref{fig:similar_modes_1}$k,m,o$ and \ref{fig:similar_modes_2}$a,g$, were identified and discussed for $\mathrm{Ra}=10^6$ and $10^8$ by \citet{soucasse2021low}. The $C^{\ast}$ and $D^{\ast}$ modes are somewhat similar to the $C$ and $D$ modes, respectively, except that each mode has rolls extending along the full height of the cell, where the rolls in their non-asterisked counterparts are confined to the upper and lower half-cells. Correspondingly, as seen in Table~\ref{tab:modes_characteristics} the $C^{\ast}$ and $D^{\ast}$ modes are antisymmetric with respect to $S_z$, while the $C$ and $D$ modes are symmetric with respect to $S_z$; symmetries are otherwise similar between the respective modes. The $BL_{x/y}^{\ast}$ modes form a pair of counter-rotating rolls (compare with the \emph{co}-rotating rolls in the $BL_{x/y}$ modes). The angular momenta of the rolls cancel out, meaning that the $BL_{x/y}$ modes as a whole carry no angular momentum. As before the $BL_{x/y}$ and $BL_{x/y}^{\ast}$ modes are linked by similar symmetries, except for symmetry vs antisymmetry with respect to $S_z$. Likewise, the $M^{\ast}$ mode is an $S_z$-antisymmetric counterpart to the $M$ mode, consisting of a single thermally unstratified toroidal roll extending the full height of the cell. Lastly, the $L_z$ mode consists of a vertical roll surrounded by horizontal rolls. This mode is the only one out of those considered capable of supporting a non-zero vertical angular momentum component, which is needed to resolve fluctuations in this component away from zero.

The last set of energy modes to be considered has not been discussed in prior literature, and consists of the modes $L_{x/y}^{\dagger}$, $M^{\dagger}$, $C^{\dagger}$, $K_{x/y}^{\dagger}$, and $D^{\dagger}$ ($\varphi_{E,13}$--$\varphi_{E,15}$ and $\varphi_{E,17}$--$\varphi_{E,20}$), shown in Figure~\ref{fig:similar_modes_2}$c,e,i,k,m$. The $L_{x/y}^{\dagger}$ modes form a degenerate pair which shares symmetry signatures with the $L_{x/y}$ and $BL_{x/y}$ pairs. The $L_x^{\dagger}$ mode features a central roll along the $x$ axis, which buckles around near the corners to form eight rolls perpendicular to the main roll, one near each corner. Together the central roll and the cross rolls give rise to temperate convection zones on the flanks of the central roll, while the cross rolls drive cold and hot convection zones along the vertical edges of the cell. The $M^{\dagger}$ mode shares its symmetry signature with the mean mode $M$. It features a pair of counter-rotating toroidal structures, localized in the lower and upper half of the cell. These produce vertically split convection zones along the vertical center line, which are surrounded by convection regions along the top and bottom edges as well as opposite convection zones towards the middle of the vertical edges. Of all modes considered here only the $M$ and $M^{\dagger}$ modes have non-vanishing thermal energy at the top and bottom walls, implying that $M^{\dagger}$ serves as a first correction to the $M$ mode in resolving the thermal stratification.

The $C^{\dagger}$ mode shares its symmetry signature with the $C$ mode, and consists of four vertical rolls along the vertical edges, separated by pairs of short horizontal rolls. It moves fluid between the vertical walls and the center of the cell. The $K^{\dagger}_{x/y}$ modes form another degenerate pair, with symmetry signatures identical to those of the $BL_{x/y}^{\ast}$ modes. The $K_x^{\dagger}$ mode consists of a horizontal roll near each of the eight corners, aligned along the $y$ axis and with alternating rotations. These are connected by an additional set of horizontal rolls lodged between the first and orthogonal to these. The angular momentum contribution of all the rolls cancel in each direction, and like the $BL_{x/y}^{\ast}$ modes the $K^{\dagger}_{x/y}$ modes therefore carry no angular momentum. Lastly, the $D^{\dagger}$ mode is unique among the modes considered in being antisymmetric with respect to all of the basic symmetry operations. It consists of eight vertical rolls aligned along the vertical sides and edges. The rolls produce eight alternatingly hot and cold convective regions, each extending the full height of the domain.

\subsection{Large scale organization of dissipation modes\label{sec:diss_modes}}
The dissipation decomposition is motivated by the desire to better resolve dynamically important structures which are not well captured by the energy decomposition. As demonstrated in Section~\ref{sec:mode_labeling} the results of the two decompositions differ both in the structures contained in the modes and in the ordering of those modes, although as evidenced by the near-unity overlaps the structural differences are mostly modest for the modes considered. In this section we take a closer look at the differences in large-scale organization between the modes, while details relating to boundary layer structures are considered in Section~\ref{sec:boundary_layer}.

Apart from reordering, the large-scale organization of the modes corresponding to $n_{E/D} \leq 12$ show little change between the two decompositions, as seen from Figures~\ref{fig:similar_modes_1} and~\ref{fig:similar_modes_2}. Several modes ($M$, $L_{x/y}$, $D$, $C$, $BL_{x/y}^{\ast}$, and $C^{\dagger}$) are found with virtually unchanged large scale organization; these modes also generally exhibit overlaps very close to unity, cf.~Table~\ref{tab:modes_characteristics}. Some changes that are notable from Figures~\ref{fig:similar_modes_1} and~\ref{fig:similar_modes_2} include a deformation of rolls in the $BL_{x/y}$ modes leading to pinching of the hot and cold convection zones at mid height and elongation of the temperate convection zones towards the top and bottom; a narrowing of rolls in the $C^{\ast}$ and $D^{\ast}$ modes also around mid height; and a splitting of the rolls at the vertical edges along the mid height plane in $M^{\ast}$.

\begin{figure*}
	\includegraphics{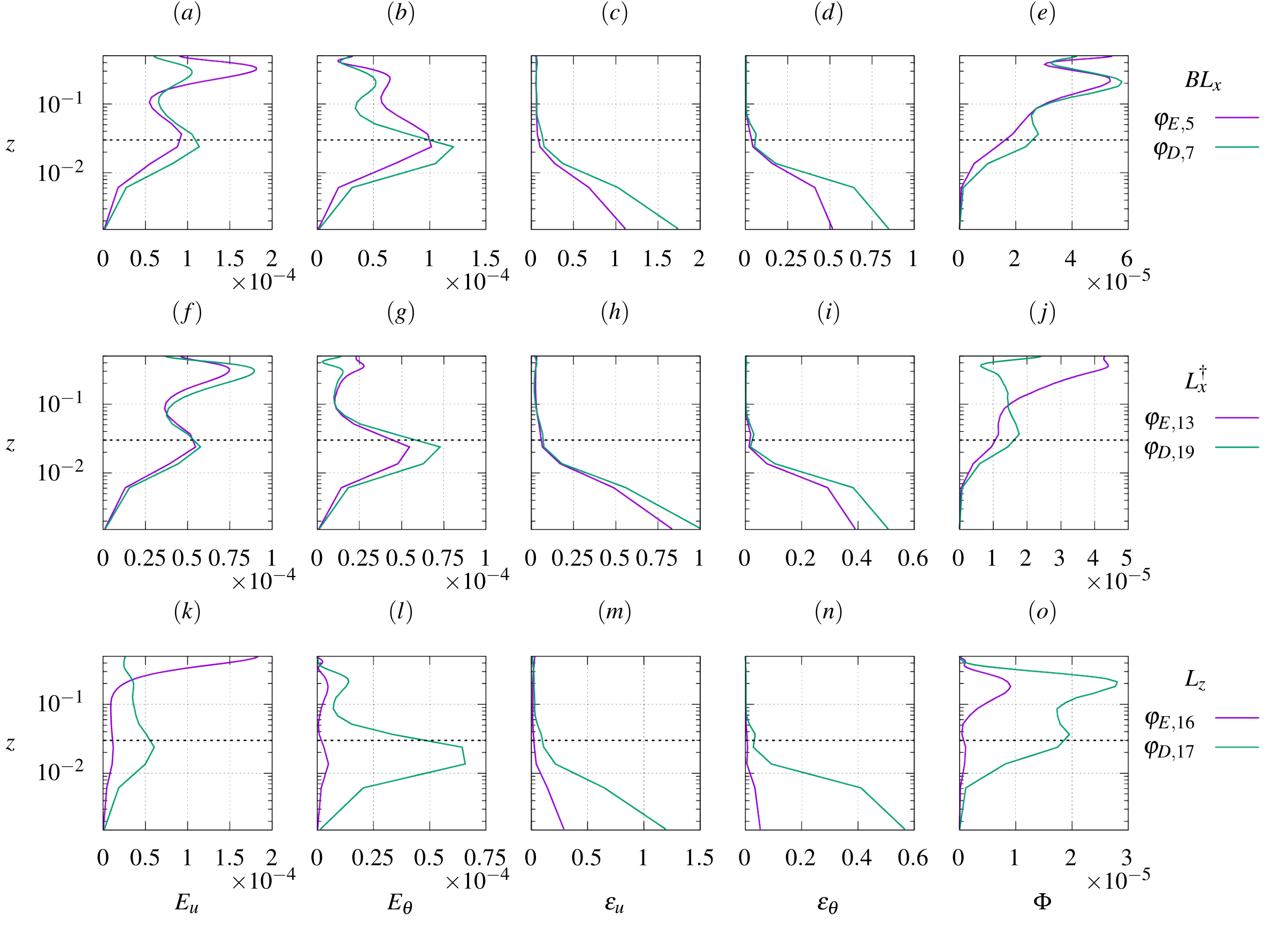}
    \caption{Profiles of kinetic and thermal energy, viscous and thermal dissipation, and convective heat flux associated with $BL_x$ modes $\varphi_{E,5}$ and $\varphi_{D,7}$ ($a$--$e$), $L_x^{\dagger}$ modes $\varphi_{E,13}$ and $\varphi_{D,19}$ ($f$--$j$), and $L_z$ modes $\varphi_{E,16}$ and $\varphi_{D,17}$ ($k$--$o$). The profiles are symmetric around the mid plane, and only $z$ values corresponding to the lower half of the cell are shown. A dotted line marks the extent of the boundary layer, $z=\delta=0.03$.\label{fig:mode_selected_profiles}}
\end{figure*}

The changes in modal structure when moving from the energy decomposition to the dissipation decomposition impact the modal contributions to reconstructed flow quantities. Vertical modal profiles of kinetic and thermal energy, viscous and thermal dissipation, and convective heat flux corresponding to the contribution of individual modes to the mean profiles are obtained by averaging individual terms in \eqref{eq:profile_convergence} along the $x$ and $y$ directions. In Figure~\ref{fig:mode_selected_profiles} we compare such profiles for some selected modes of each decomposition. The profiles for the $M$ and $L_{x/y}$ modes (not shown) are virtually identical between decompositions. For the $D$, $BL_{x/y}$, and $C$ modes the profiles are very similar among the decompositions, showing only minor variations; the greatest deviation for those modes is found for the $BL_x$ mode profiles, which are shown in Figure~\ref{fig:mode_selected_profiles}$a$-$e$. For higher modes the profiles differ between decompositions in a mostly consistent manner, with notable exceptions highlighted below. While the shape of profiles varies somewhat between modes, for each of the quantities considered the profiles obtained from dissipation modes generally contribute more than the energy modes in the region near the top and bottom. The contributions are reversed or similar in the core of the cell.

In the $L_x^{\dagger}$ dissipation mode in Figure~\ref{fig:similar_modes_2}$d$ the central roll is widened compared to the energy mode in Figure~\ref{fig:similar_modes_2}$c$, and the cross rolls align with the horizontal diagonals instead of being orthogonal to the roll axis. This also causes a wider temperate convection zone, splitting the edge convection zones vertically. The diagonally aligned cross rolls produce additional convection zones above and below the central roll. The splitting of the edge convection zone diminishes the thermal fluctuation and the heat flux near mid-height of the cell, resulting in a reduced contribution from the dissipation mode to the vertical mean thermal energy and convective heat flux profiles shown in Figure~\ref{fig:mode_selected_profiles}$g,j$. The substantially different structures and heat flux profiles found for the energy and dissipation $L_{x/y}^{\dagger}$ modes ($\varphi_{E,13}$ and $\varphi_{D,19}$) are reflected in the rather modest overlap between the modes, at $\Gamma(13,19)=0.6037$.

In the $M^{\dagger}$ dissipation mode in Figure~\ref{fig:similar_modes_2}$f$ the opposite convection zones near the corners are strengthened compared to the energy mode, and they extend further towards the horizontal walls and mid plane of the domain. The convection zone along the top and bottom edges largely disappear.

The $L_z$ modes shown in Figure~\ref{fig:similar_modes_2}$g,h$ exhibit substantial dissimilarities between the decompositions. In the dissipation $L_z$ mode the vertical roll is much narrower than in the energy mode, and the mode is instead dominated by the surrounding horizontal rolls. The dissimilarity is also supported by the mode overlaps; while the greatest overlap for the energy $L_z$ mode ($\varphi_{E,16}$) found among the modes investigated here was with the $L_z$ dissipation mode ($\varphi_{D,17}$) and vice versa, at $\Gamma(16,17) = 0.6461$, both exhibit slightly greater overlaps with different modes of higher indices; in particular, $\Gamma(30, 17) = 0.7053$, and $\Gamma(16, 72) = 0.7066$. For all other modes considered here the maximum overlapping mode was found within the set $n_{E/D}\leq 20$. The dissimilarity between the $L_z$ modes is also found in the vertical profiles in Figure~\ref{fig:mode_selected_profiles}$k$--$o$, which differ rather substantially between the two modes. The kinetic energy profile of the $L_z$ energy mode has a marked peak at mid-height, which is entirely absent in the $L_z$ dissipation mode. Conversely, the dissipation mode thermal energy profile peaks within the boundary layer, while the corresponding contribution from the energy mode is much smaller throughout the domain. This suggests that the energy mode is mainly kinetic in nature, while the dissipation mode is mainly thermal. The dissipation mode contribution to both viscous and thermal dissipation near the wall is notably stronger than that of the energy mode, which agrees qualitatively with the trends displayed by other modes. Finally, due to the diminished thermal fluctuations in the energy mode, the dissipation mode contributes significantly more to convective heat flux throughout the cell (except at mid height) than does the energy mode.

While the primary rolls in the $K_{x/y}^{\dagger}$ energy mode (Figure~\ref{fig:similar_modes_2}$k$) are aligned along the $y/x$ axis and joined by orthogonal horizontal rolls, for the corresponding dissipation mode (Figure~\ref{fig:similar_modes_2}$l$) the first set of rolls are instead aligned along the horizontal diagonals, and cross-rolls are notably wider. In the $D^{\dagger}$ dissipation mode (Figure~\ref{fig:similar_modes_2}$n$) the rolls along the vertical edges are split into pairs of co-rotating rolls.

To summarize, there is generally a large degree of similarity in the large-scale organization between analogous modes of different decompositions. This is especially striking for the lower modes in Figure~\ref{fig:similar_modes_1}, whereas the higher modes in Figure~\ref{fig:similar_modes_2} tend be more distinct, in agreement with the overlaps summarized in Table~\ref{tab:modes_characteristics}. Part of the explanation for these similarities is the symmetry-imposed constraints on mode mixing. The two bases produced by the decompositions span the same Hilbert space, indicating that each mode of one basis can be expanded as a linear combination of modes from the other basis. As was mentioned in Section~\ref{sec:mode_labeling}, the overlap vanishes for pairs of modes which do not share their configuration of symmetry eigenvalues, implying that only isosymmetric modes enter in this expansion. This prohibits mixing between different families of isosymmetric modes, constraining the dissimilarity between modes produced by different decompositions.

Even within such families we find mixing to be limited, however. Among all modes considered the $L_{x/y}^{\dagger}$ modes have the smallest overlap between analogous modes, with $\Gamma(13,19) = 0.6037$ for the $L_x^{\dagger}$ modes $\varphi_{E,13}$ and $\varphi_{D,19}$. The modes exhibit clear differences in large-scale structures between the decompositions, as shown in Figure~\ref{fig:similar_modes_2}$c,d$. This dissimilarity is due to mixing within the isosymmetric family $\{L_{x/y}, BL_{x/y}, L_{x/y}^{\dagger}, \ldots\}$, including a substantial overlap of $\Gamma(5,19) = 0.5828$ between the $BL_x$ mode $\varphi_{E,5}$ and the $L_x^{\dagger}$ mode $\varphi_{D,19}$. On the other hand, little mixing happens between the $L_{x/y}$ and $BL_{x/y}$ modes, with $\Gamma(2,7) = 0.1255$ and $\Gamma(5,2) = 0.0211$. Other isosymmetric families include $\{M,M^{\dagger},\ldots\}$, $\{C,C^{\dagger},\ldots\}$, and $\{BL_{x/y}^{\ast}, K_{x/y}^{\dagger},\ldots\}$; naturally, the remaining modes also belong to isosymmetric families, although those modes are each the sole representative of their respective families within the set of modes considered here. Mixing within the families considered here is generally rather limited, which results in large-scale organization being mostly stable across decompositions. Again, however, we remark that while this is true for modes of low indices, the stability diminishes when considering higher index modes.

The stability up to reordering of large-scale organization of modes under varying decompositions was also found by \citet{podvin2015large} when comparing kinetic and thermal energy modes for a two-dimensional square Rayleigh-Bénard cell. They attributed the similarity between the decompositions to the close coupling between velocity and temperature, as the motion is entirely driven by temperature fluctuations. In the present work, the observation that the first several modes of both energy and dissipation decompositions describe structures much more similar than what is required by symmetry constraints supports the hypothesis of a coupling between energetic and dissipative structures discussed in Section~\ref{sec:mode_labeling}. Due to the modest Rayleigh and Reynolds numbers $\mathrm{Ra}=10^7$ and $\mathrm{Re} = 651$, as defined in \eqref{eq:rayleigh} and \eqref{eq:reynolds}, we expect only a moderate separation between energetic and dissipative scales, leading to a tighter coupling between the associated structures. We would thus expect a greater amount of mode dissimilarity and reordering given higher values of $\mathrm{Ra}$ and $\mathrm{Re}$.

We note that the remarkable stability of the LSC modes $M$, $L_{x/y}$, and $D$ across decompositions may further be attributed to the degree to which the structures they encode dominate the flow.

The stability of large-scale structures between decompositions notwithstanding, certain subtle but consistent changes are found. Dissipation modes tend to contribute more to vertical mean profiles of kinetic and thermal energy, viscous and thermal dissipation, and heat flux in the boundary layer region than do energy modes. This trend is seen in Figure~\ref{fig:mode_selected_profiles}, and it is also found in the corresponding profiles of other modes (not shown). Conversely, in the core of the domain the contributions are either similar between decompositions or diminished for dissipation modes. A similar observation was made by \citet{olesen2023dissipation} for profiles of kinetic energy and of viscous dissipation reconstructed using decompositions optimized with respect to these quantities. This was ascribed to the different spatial distributions of the quantities with respect to which the decompositions were optimized. Kinetic energy is located mainly in the core of the cell, viscous dissipation is concentrated along the walls (both horizontal and vertical), and thermal energy and thermal dissipation peak near the top and bottom. The energy and dissipation decompositions each give equal weight to the two terms of energy and dissipation, respectively. This causes the energy decomposition to give roughly equal emphasis to the core and boundary regions, and the dissipation decomposition to emphasize structures in the boundary layer region.

Finally, we note that Figure~\ref{fig:mode_selected_profiles} demonstrates the failure of the assumption used in isotropic eddy-diffusivity models, as pointed out by \citet{hanjalic2002one}, that heat flux locally scales as $\Phi\sim \sqrt{\left\langle E_u E_{\theta}\right\rangle}$. For example, in the energy $BL_{x/y}$ mode both $E_u$ and $E_{\theta}$ show local minima at $z=0.1$ (Figure~\ref{fig:mode_selected_profiles}$a,b$), while $\Phi$ increases monotonically with $z$ up to around $z=0.2$ (Figure~\ref{fig:mode_selected_profiles}$e$). This is attributed to the unequal distribution of kinetic energy among the available degrees of freedom.

\subsection{Boundary layer structures\label{sec:boundary_layer}}
As discussed in Section~\ref{sec:introduction}, boundary layers are of special interest in Rayleigh-Bénard convection. In this section we consider the boundary layer structures resolved by energy and dissipation modes. Since viscous and thermal dissipation are both concentrated in the boundary layer region we expect this region to be particularly sensitive to the choice of decomposition. The analysis presented here highlights some subtle but potentially important differences between the decompositions.

\begin{figure}
	\includegraphics{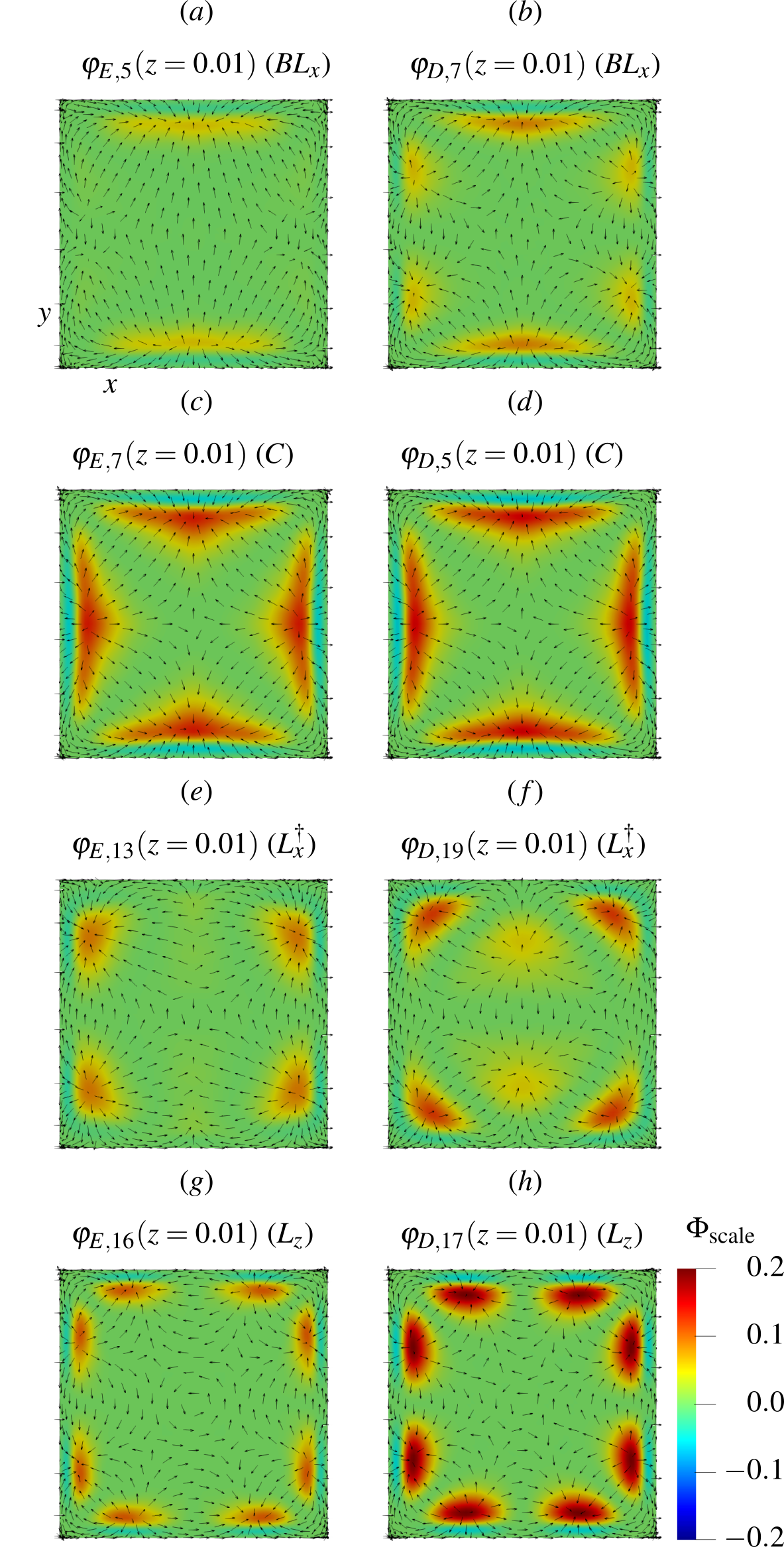}
    \caption{Horizontal cross sections taken near the bottom ($z=0.01$) of the $BL_x$ modes $\varphi_{E,5}$ ($a$) and $\varphi_{D,7}$ ($b$), the $C$ modes $\varphi_{E,7}$ ($c$) and $\varphi_{D,5}$ ($d$), the $L_x^{\dagger}$ modes $\varphi_{E,13}$ ($e$) and $\varphi_{D,19}$ ($f$), and the $L_z$ modes $\varphi_{E,16}$ and $\varphi_{D,17}$. Slices are colored by scaled convective heat flux, $\Phi_{\mathrm{scale}} = \varphi^w\varphi^{\theta}/\max_{\Omega}\left({\varphi^w\varphi^{\theta}}\right)$. Arrows show the direction of the in-plane velocity.
    \label{fig:slice_selected}}
\end{figure}

In Figure~\ref{fig:slice_selected} we show horizontal cross sections of the bottom boundary layer ($z=0.01$) of selected modes. The cross sections are colored by heat flux, and also show the direction of the horizontal components of the flow field. The LSC modes $M$, $L_{x/y}$, and $D$ were found in Section~\ref{sec:diss_modes} to be largely stable under the change of decomposition, and the cross sections obtained from these modes (not shown) exhibit no discernible differences between the decompositions. For the $BL_x$ modes shown in Figure~\ref{fig:slice_selected}$a,b$ the dissipation mode exhibits an enhanced heat flux along the edges at $y=0$ and $y=1$, as well as in separated pairs of patches along $x=0$ and $x=1$. These correspond to the convective regions seen in Figure~\ref{fig:similar_modes_1}$g,h$, reflecting the extension of the temperate convective regions discussed in Section~\ref{sec:diss_modes}. Figure~\ref{fig:slice_selected}$a,b$ thus demonstrates that the changes in these features extend well into the boundary layer. This extension and strengthening of features into the boundary layer in dissipation modes compared to analogous energy modes represents a general trend that is also seen for the $L_x^{\dagger}$ and $L_z$ modes shown in Figure~\ref{fig:slice_selected}$e$--$h$, as well as for most of the modes $n_{E,D}\leq 20$ not shown in Figure~\ref{fig:slice_selected}.

A deviation from this trend is seen in the horizontal cross sections of the bottom boundary layer of the $C$ modes shown in Figure~\ref{fig:slice_selected}$c,d$. We note little difference between the two decompositions, although a slight narrowing of the wedge-shaped regions of positive (upward) heat flux can be discerned. The cross sections show a high level of activity in the boundary layer for both modes, consisting of said wedge-shaped zones marginally separated from narrow stretches of slight negative (downward) heat flux along the edges. Comparing with Figure~\ref{fig:similar_modes_1}$i,j$ we see that the wedge-shaped zones correspond to downwards movement of cold fluid near the $x=0$ and $x=1$ walls, and upwards movement of warm fluid near the $y=0$ and $y=1$ walls. The zones of negative heat flux are found to correspond to inversions of the vertical velocity, implying large velocity gradients in the margins separating the regions. It was noted in Section~\ref{sec:mode_labeling} that the $C$ mode is important in promoting LSC reorientations, and that its promotion in the dissipation decomposition could reflect an improved representation of the flow dynamics in this decomposition; in this light it is interesting to note that the mode itself appears to change relatively little in its large-scale organization as well as in its boundary layer layout between the decompositions.

The $L_{x/y}^{\dagger}$ and $L_z$ modes were found in Section~\ref{sec:diss_modes} to differ substantially in their large scale organization between the decompositions. Horizontal cross sections of the $L_x^{\dagger}$ modes are shown in Figure~\ref{fig:slice_selected}$e,f$. Moving from the energy to dissipation decomposition brings about changes in the cross section reminiscent of those seen for $BL_x$ in Figure~\ref{fig:slice_selected}$a,b$, with regions of positive heat flux towards the corners being stronger and more well-defined in the dissipation mode, as well as additional regions appearing around $x=0.5$. This corresponds to the additional regions of positive convection appearing in Figure~\ref{fig:similar_modes_2}$d$ compared to Figure~\ref{fig:similar_modes_2}$c$ as discussed in Section~\ref{sec:diss_modes}.

We show horizontal cross sections of the $L_z$ mode in Figure~\ref{fig:slice_selected}$g,h$. While the qualitative pattern observed changes little between the decompositions, the convective regions are substantially enlarged and strengthened in the dissipation mode. Comparing with Figure~\ref{fig:similar_modes_2}$g,h$ we see that the stronger vertical movement in the dissipation mode compared to the energy mode causes the convective regions to reach further towards the top and bottom.

To summarize, there is an general trend towards increased boundary layer activity in the dissipation modes compared to energy modes, expressed through more extensive and well-defined zones of substantial convective heat flux in these modes.

\begin{figure}
	\includegraphics{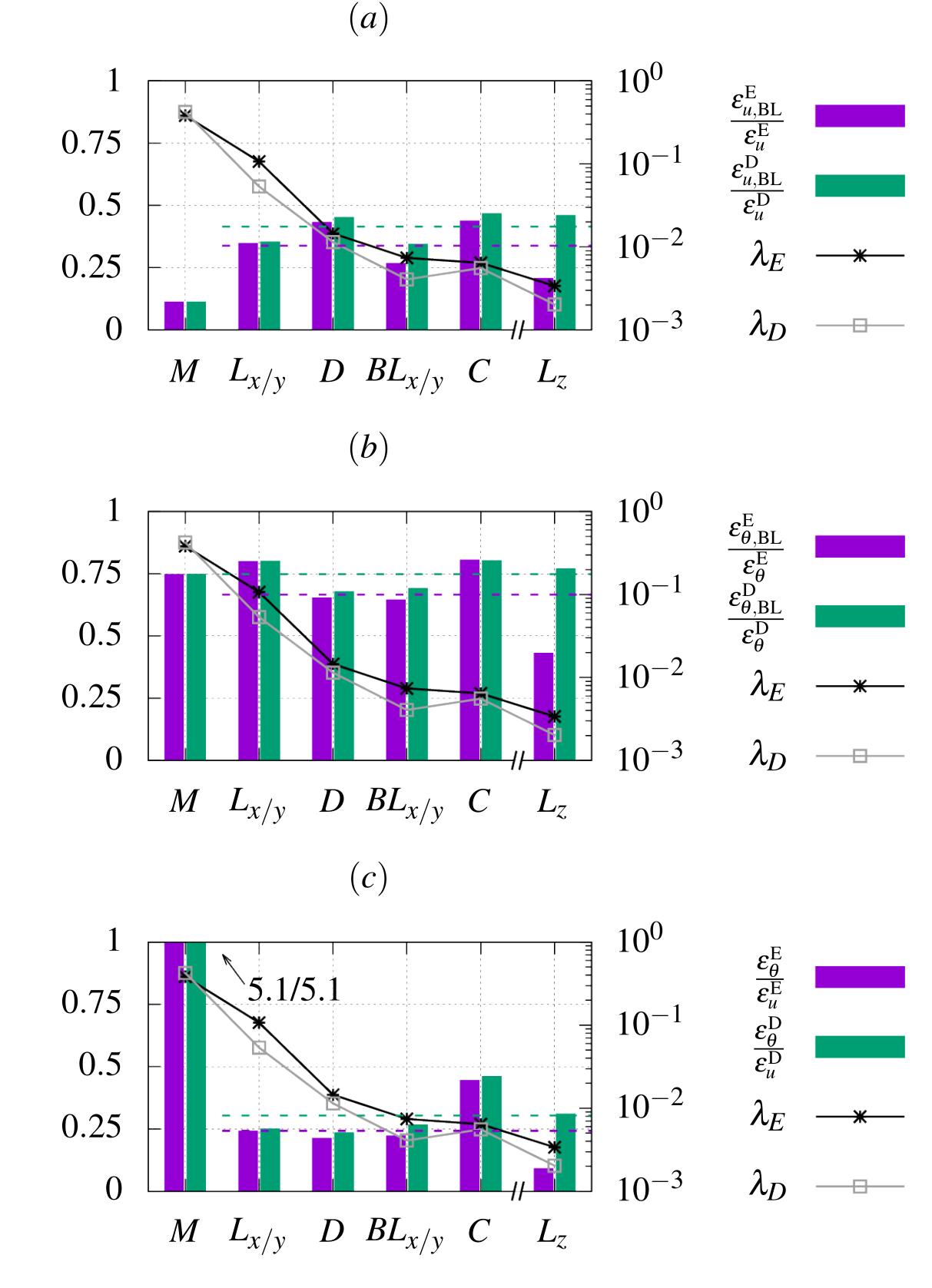}
    \caption{Modal viscous $(a)$ and thermal $(b)$ dissipation fraction in boundary layer, and modal ratio  of thermal to viscous dissipation $(c)$ for the modes up to $C$ (left axes). Dashed lines show the mean values for each decomposition, excluding the $M$ modes. The relative importance of modes is given by eigenvalues, which are plotted with lines (right axes).
    \label{fig:bl_diss}}
\end{figure}

The theory proposed by \citet{grossmann2000scaling,grossmann2004fluctuations} referred to in Section~\ref{sec:introduction} suggests that important modeling parameters include the relative contributions of viscous and thermal dissipation in the boundary layers and in the whole cell. This motivates analyzing the modes in terms of these parameters. In Figure~\ref{fig:bl_diss} we compare on a per-mode basis the ratio of boundary layer contributions to the full-cell viscous (Figure~\ref{fig:bl_diss}$a$) and thermal dissipation (Figure~\ref{fig:bl_diss}$b$) for energy and dissipation modes, as well as the ratio of thermal to viscous dissipation (Figure~\ref{fig:bl_diss}$c$), for the LSC modes $M$, $L_{x/y}$, and $D$; the $BL_{x/y}$ and $C$ modes describing the leading fluctuation terms; and the $L_z$ mode which was found to deviate from several of the trends established in Sections~\ref{sec:diss_modes} and~\ref{sec:boundary_layer}. The values averaged over the modes (excluding the $M$ modes) are shown by dashed lines, and eigenvalues are shown to indicate the relative importance of the modes in the reconstructed flow.

Figure~\ref{fig:bl_diss}$a$ shows the fraction of total viscous dissipation located in the top and bottom boundary layers. These contributions amount to 0.2--0.5, except for the $M$ mode where it is around 0.1. Averaged over modes $1<n_{E/D} \leq 20$ (i.e., excluding $M$ modes) the mean fraction is 0.32 for energy  modes and 0.41 for dissipation modes. The boundary layer contributions are enhanced for most dissipative modes compared to energy modes, except for the $M$ and $L_{x/y}$ modes where ratios are nearly identical; a more substantial difference is observed for the $L_z$ modes. The trend of enhanced boundary layer contributions for dissipation modes is maintained by the remaining modes $n_{E/D}\leq 20$ (not shown), with the difference between the fractions for dissipation modes and for energy modes tending to increase for higher modes. The enhanced boundary layer contributions from dissipation modes reflect the increased boundary layer activity that was found for most of the modes considered in Figure~\ref{fig:slice_selected}. 

The fraction of thermal dissipation contributed from the boundary layer is shown in Figure~\ref{fig:bl_diss}$b$, where we also find approximately constant ratios around 0.6--0.8 among the modes considered, with the exception of the $L_z$ energy mode for which the fraction is around 0.4. The average fractions for modes $1<n_{E/D}\leq 20$ are 0.69 for energy modes and 0.76 for dissipation modes. The thermal dissipation boundary layer fraction is thus generally on the order of twice the viscous dissipation boundary layer fraction. This difference is likely caused by the exclusion of the vertical boundary layers, which contribute appreciable viscous dissipation and little thermal dissipation. The difference in thermal dissipation boundary layer fraction between energy and dissipation modes is generally small compared to the difference in the viscous dissipation boundary layer fraction, with the most substantial difference again found for the $L_z$ modes. The increased boundary layer activity for dissipation modes compared to energy modes thus has a smaller relative impact on the thermal dissipation boundary layer fraction than on the viscous dissipation boundary layer fraction. 

The ratios of thermal to viscous dissipation over the whole cell are shown in Figure~\ref{fig:bl_diss}$c$. The $M$ modes stand out with a ratio of 5.1, their dissipation being strongly dominated by the thermal contribution. This supports the finding by \citet{soucasse2019proper} that the $M$ modes capture the thermal stratification of the flow. The remaining modes (including the higher modes not shown here) are dominated by viscous dissipation, with average ratios for modes $1<n_{E/D}\leq 20$ of 0.26 for energy modes and 0.34 for dissipation modes. For the modes up to and including $C$ ratios are mostly consistent between energy and dissipation modes. Variability between the decompositions increases for higher modes, in agreement with the trend of decreasing similarity found in Sections~\ref{sec:mode_labeling} and~\ref{sec:diss_modes}. The $C$ modes, along with a few higher modes, have substantially higher ratios than those found for $L_{x/y}$, $D$, and $BL_{x/y}$ modes. It is possible that the destabilizing effect of the $C$ modes on the LSC is linked to the relative importance of thermal dissipation in this mode. The $L_z$ energy mode has a remarkably small ratio, whereas the ratio of the corresponding dissipation mode matches the average dissipation mode ratio. This matches the observation from Figure~\ref{fig:mode_selected_profiles} in Section~\ref{sec:diss_modes} that the energy mode was mainly kinetic in nature, whereas the dissipation mode was mainly thermal.

We see from the preceding analysis that dissipation modes in general tend to have a greater fraction of both viscous and thermal dissipation localized in the boundary layers than do energy modes. This is a consequence of the increased activity in the boundary layer for  dissipation modes. Also, while modes of either decomposition are as a rule dominated by viscous dissipation, dissipation modes tend toward a higher ratio of thermal to viscous dissipation. 

\section{Modal convergence\label{sec:convergence}}
The energy and dissipation POD bases each allow reconstruction of second order mean quantities using \eqref{eq:profile_convergence}. Here we investigate the convergence of such reconstructions of kinetic and thermal energy, viscous and thermal dissipation, and convective heat flux. We consider the convergence of quantities integrated over the full domain ($\Omega$) as well as over the region consisting of the horizontal top and bottom boundary layers, $BL=[0;1]\times [0;1]\times \left([0;\delta]\cup[1-\delta;1]\right)$,
\begin{align}
    \Pi^{p,\Omega'}(n) &= \int_{\Omega'} \Pi^p_n\, dx\,,
\end{align}
where $\Omega'\in\left\{\Omega, BL\right\}$. Here we use $\delta=0.03$ as an estimate of the boundary thickness, based on the kinetic and thermal boundary layer thicknesses $\delta_u=0.027$ and $\delta_{\theta} = 0.031$\citep{shishkina2010boundary}.

\begin{figure*}
	\includegraphics{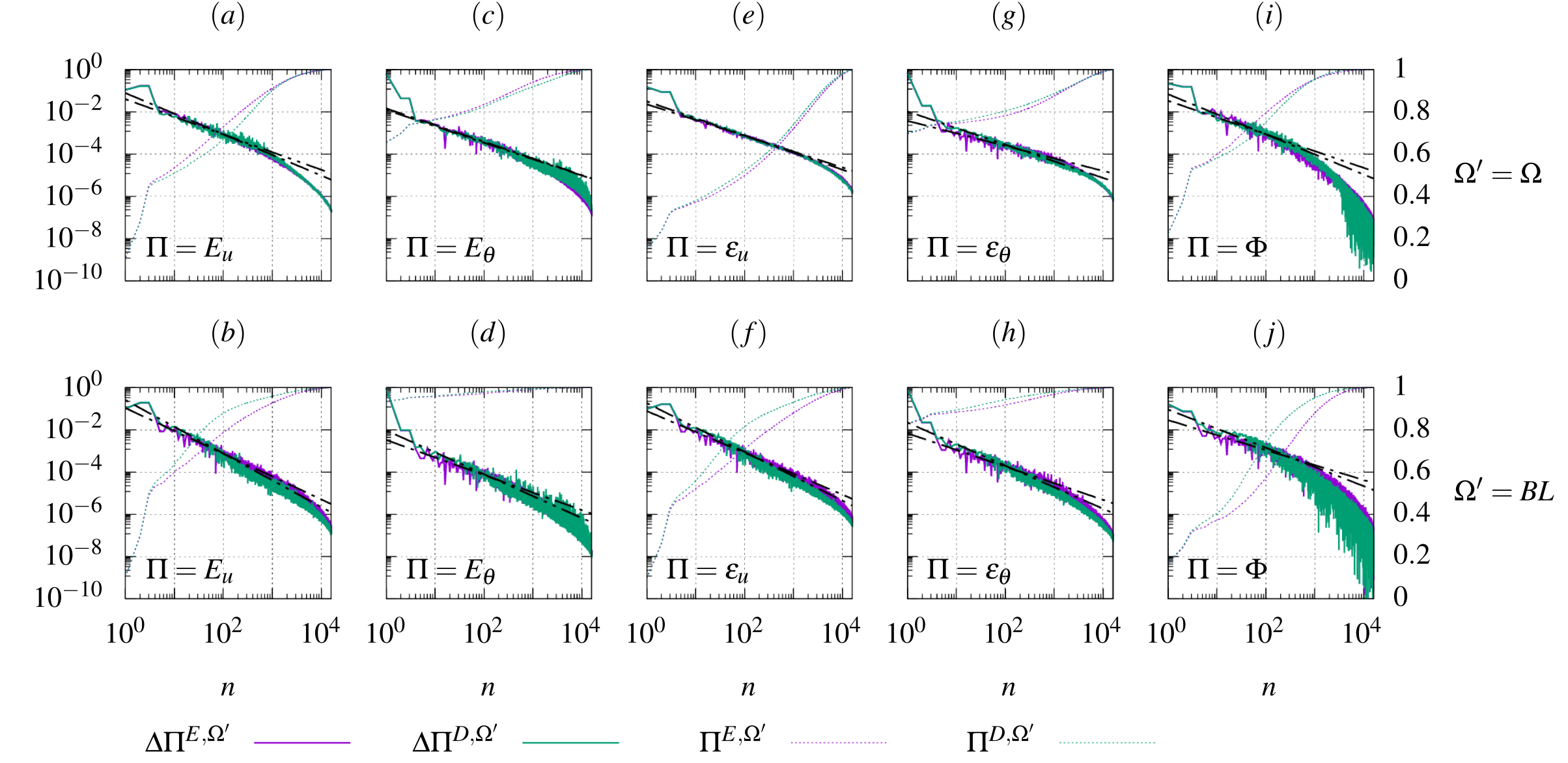}
    \caption{Modal convergence of kinetic ($a, b$) and thermal energy ($c, d$), viscous ($e, f$) and thermal dissipation ($g, h$), and heat flux ($i, j$) in the full domain (top row) and in the boundary layer ($z\in[0;\delta]\cup[1-\delta;1]$ with $\delta=0.03$, bottom row). Fully drawn lines show the per-mode contribution (left axes), and dotted lines the cumulative value (right axes). Also shown are power laws fitted to per-mode contributions in the interval $10\leq n\leq 10^3$. All values are normalized with respect to their full reconstruction.
    \label{fig:convergence}}
\end{figure*}

We compare the convergence achieved for each integrated quantity using the energy decomposition to that achieved with the dissipation decomposition. The convergence of each reconstruction is shown in Figure~\ref{fig:convergence}. Considering the cumulative values (dotted lines) we see that for all reconstructed quantities the three lowest energy modes contribute the same as the corresponding dissipation modes, as these modes are nearly identical (cf.\ Section~\ref{sec:pod_specta_modes}). The $M$ mode captures much of the thermal stratification in the flow, and accordingly accounts for most of the thermal energy and thermal dissipation in both the full domain and in the boundary layers.

For the reconstruction of kinetic energy $E_u$ in the full domain, shown in Figure~\ref{fig:convergence}$a$, the energy decomposition converges slightly faster than the dissipation decomposition, as would be expected because of the energy optimization. However, the converse is true for the reconstruction of kinetic energy in the boundary layer (Figure~\ref{fig:convergence}$b$). The convergence of thermal energy (Figure~\ref{fig:convergence}$c,d$) is very rapid, and the two decompositions show little difference in convergence rates, in the full domain as well as in the boundary layer. A very small lead can be ascertained for the energy decomposition in the full domain.

For the reconstructions of viscous ($\varepsilon_u$) and thermal dissipation ($\varepsilon_{\theta}$) in the full domain (Figure~\ref{fig:convergence}$e,g$) the two decompositions are again very similar, with a marginal advantage to the dissipation decomposition; the same applies to the reconstruction of the thermal dissipation in the boundary layer (Figure~\ref{fig:convergence}$h$). However, in the reconstruction of viscous dissipation in the boundary layer (Figure~\ref{fig:convergence}$f$) the dissipation decomposition exhibits a substantially more rapid convergence than the energy decomposition.

The reconstruction of convective heat flux $\Phi$ in the full domain and in the boundary layer (Figure~\ref{fig:convergence}$i,j$) repeat the pattern for kinetic energy, with faster convergence using the energy decomposition in the full domain, and using the dissipation decomposition in the boundary layer. However, given the very limited amount of convective heat flux in the boundary layer the convergence rate of convective heat flux in the boundary layer is of limited importance.

The overall picture emerging from this is that in the boundary layer the dissipation decomposition reconstructs all quantities investigated at least as efficiently as the energy decomposition, while results are more mixed when considering the full domain. This suggests that the dissipation decomposition not only gives the most efficient representation of total dissipation, for which it is optimized, but provides an overall improvement in the spectral resolution of the boundary layer region.

The per-mode contributions for each quantity in Figure~\ref{fig:convergence} approximately follow power law decays in the range \hbox{$10\leq n \leq 10^3$}. The exponents have been determined by fitting power functions to per-mode contributions in this interval, and are summarized in Table~\ref{tab:convergence_exponents}. In agreement with the trends identified above, the energy decomposition for the full cell exhibits faster convergence for kinetic energy and convective heat flux compared to the dissipation decomposition, while the convergence rates are nearly identical for thermal energy. Conversely, the dissipation decomposition gives the faster convergence rate for both viscous and thermal dissipation. In the boundary layer the dissipation decomposition yields faster convergence for all quantities than does the energy decomposition. This also includes thermal energy, for which no substantial difference on convergence rate was apparent from Figure~\ref{fig:convergence}$d$ due to 
the very fast initial convergence caused by the thermally stratified $M$ mode. We observe that all quantities converge faster in the boundary layer than in the full domain, with the exception of the energy reconstruction of the convective heat flux, although as mentioned the convective heat flux contribution from the boundary layer is negligible. Finally, we note that the dissipation-based reconstructions in the full cell converge with approximately the same rate ($\alpha\approx -0.8$) for all quantities. This might be taken as an indication that the dissipation decomposition is able to capture a wide range of scales more uniformly than is the case for the energy decomposition.

\begin{table}[ht]
    \centering
    \begin{tabular}{cc|c|c|c|c|c}
        $\Omega'$ & $p$ & $E_u$ & $E_{\theta}$ & $\varepsilon_u$ & $\varepsilon_{\theta}$ & $\Phi$ \\ \hline
         \multirow{2}{*}{$\Omega$} & $E$ & -0.98 & -0.79 & -0.75 & -0.59 & -0.95 \\
         & $D$ & -0.84 & -0.78 & -0.81 & -0.79 & -0.79 \\ \hline
         \multirow{2}{*}{$BL$} & $E$ & -1.08 & -0.83 & -0.99 & -0.79 & -0.71 \\
         & $D$ & -1.27 & -1.04 & -1.15 & -1.02 & -0.91
    \end{tabular}
    \caption{Asymptotic decay exponents $\alpha$ for the convergence plots shown in Figure~\ref{fig:convergence}. The exponents are obtained by fitting power functions $\Delta\Pi^{p,\Omega'}\sim n_p^{\alpha}$ in the range $10\leq n \leq 10^3$ to per-mode contributions to each reconstruction.}
    \label{tab:convergence_exponents}
\end{table}

\begin{figure*}
	\includegraphics{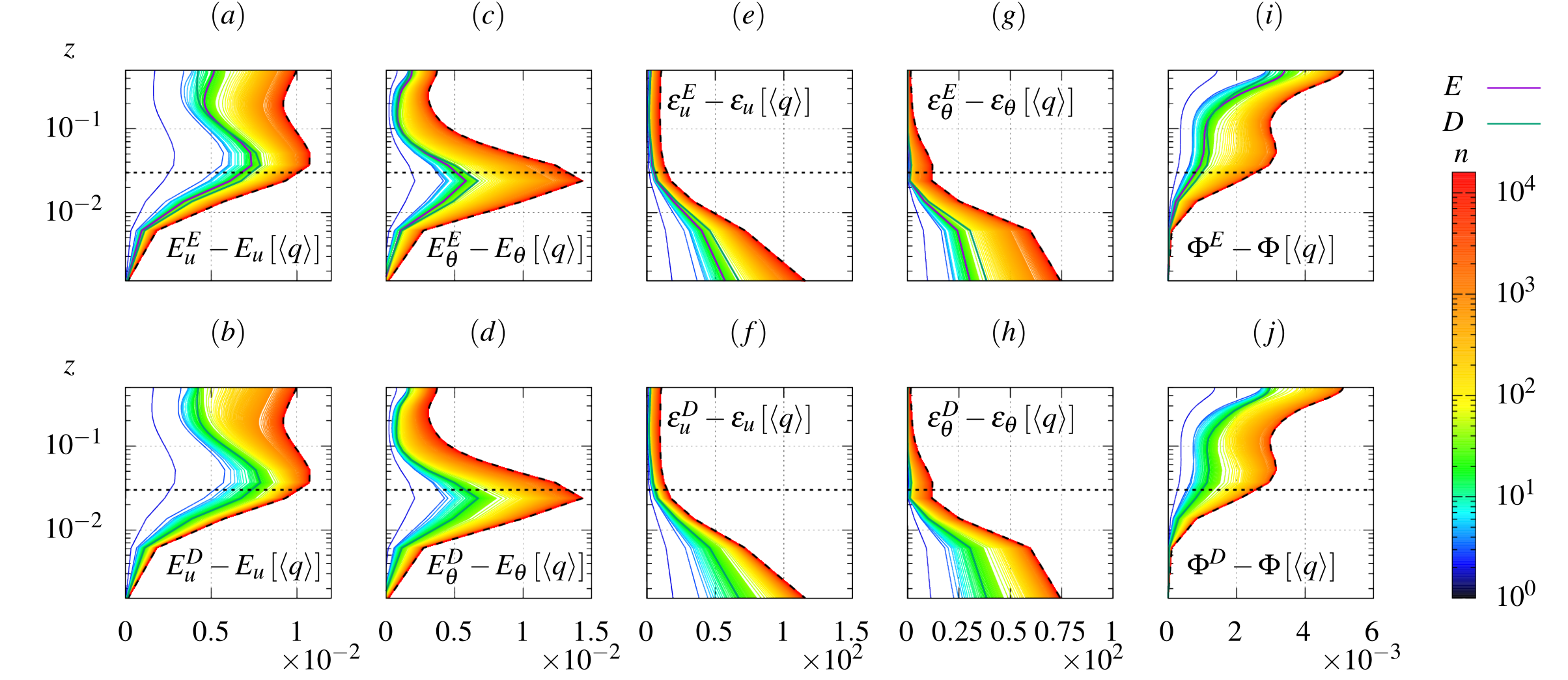}
    \caption{Cumulative reconstructions of vertical profiles, with the mean field profile $\Pi\left[\left\langle q\right\rangle\right]$ subtracted, of kinetic energy ($a,b$), thermal energy ($c,d$), viscous dissipation ($e,f$), thermal dissipation ($g,h$), and convective heat flux ($i,j$) reconstructed using energy (top row) and dissipation decomposition (bottom row). Shown are each profile using the first $n$ modes for each $n \leq 40$, every 10 profiles for $40<n\leq800$, and every 100 profiles for $800<n<\num{16000}$. Colors show the number of modes included. The mean profile is shown as a dashed line, and a dotted line at $z=\delta=0.03$ marks the extent of the boundary layer. Profiles corresponding to $n=20$ are emphasized with darker lines, cf.~the legend on the upper right. The profiles for $n=20$ from the dissipation decomposition (lower row plots) are reproduced in the upper row plots for comparison. All profiles are vertically symmetric and are shown only for the lower half of the domain.
     \label{fig:cumulative_profiles}}
\end{figure*}

We may also reconstruct full profiles using \eqref{eq:profile_convergence}. Figure~\ref{fig:cumulative_profiles} shows the mode-by-mode cumulative reconstruction of vertical profiles (with the mean field profile subtracted) of kinetic and thermal energy, viscous and thermal dissipation, and convective heat flux. In the figure, the color of each reconstructed profile corresponds to the number of modes included for that profile, and differences in the shape of each isochrome between the respective decompositions signify local differences in convergence of the profile. Note that each full reconstruction ($n=N$) exactly matches the corresponding mean profile, illustrating the completeness of both bases.

As discussed above the lowest four modes are nearly identical, and the profiles corresponding to modes 2, 3, and 4 in Figure~\ref{fig:cumulative_profiles} differ only imperceptibly between the decompositions. The profiles corresponding to mode 1 are nearly identical to the subtracted mean field profile, and are not shown. For higher modes differences can be ascertained by studying the details in each plot. When a given isochrome extends further to the right in the energy decomposition profile than in the corresponding dissipation decomposition profile for a given vertical position it implies a faster convergence of the energy decomposition at that position, and vice versa. 

However, extracting meaningful differences in the reconstruction by inspecting isochromes in Figure~\ref{fig:cumulative_profiles} can be difficult. Therefore, profiles corresponding to $n=20$ have been marked in 
Figure~\ref{fig:cumulative_profiles}, and the corresponding profiles from the two decompositions are shown together for comparison in the top row plots in Figure~\ref{fig:cumulative_profiles}. Comparing the emphasized reconstructions for $n=20$ allows us to see the variations in convergence with the vertical coordinate more clearly for these particular reconstructions. The energy modes reconstruct more of the emphasized energy and heat flux profiles near the center of the cell, while little difference can be seen for the remaining quantities. The converse is true near the lower wall, with more of each profile being reconstructed using the dissipation decomposition than using the energy decomposition.

In agreement with what was found from the global convergence discussed above, we find a trend of greater reconstruction efficiency near the boundary of the domain using dissipation decomposition, and near the core of the domain using the energy decomposition.

\begin{figure*}
	\includegraphics{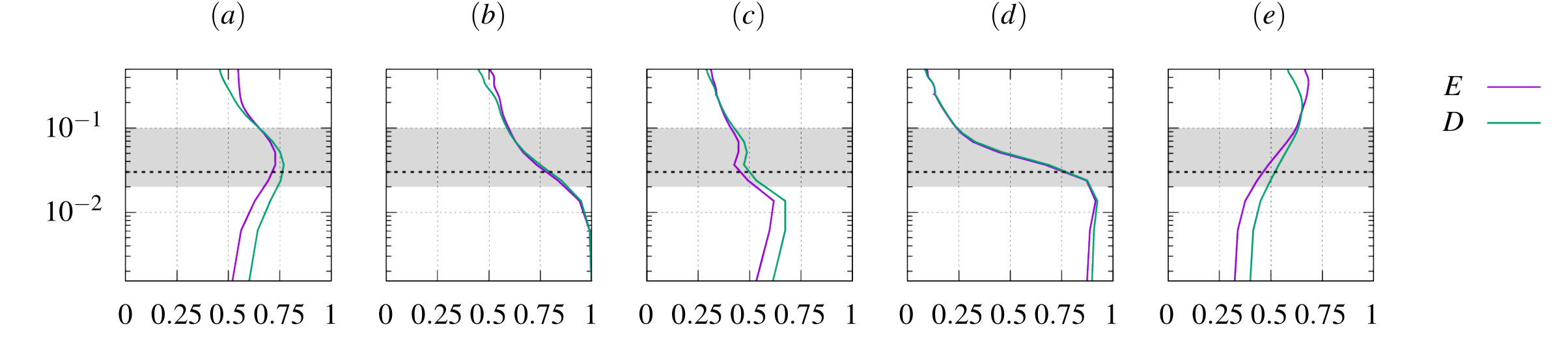}
    \caption{Ratio of reconstructed profiles using $n=20$ modes to the full mean profile, $\frac{\Pi^p_{n}}{\left\langle\Pi\right\rangle}$ for $\Pi=E_u$ ($a$), $E_{\theta}$ ($b$), $\varepsilon_{u}$ ($c$), $\varepsilon_{\theta}$ ($d$), and $\Phi$ ($d$), and $p\in{E,D}$. A dotted line at $z=\delta=0.03$ marks the extent of the boundary layer. Gray regions mark the approximate extent of the entrainment zone, $0.02\lesssim z \lesssim 0.1$. All profiles are symmetric and shown only for the lower half of the cell.
    \label{fig:scaling}
    }
\end{figure*}

\section{Perspectives for modeling\label{sec:modeling_perspectives}}
An important motivation for considering the dissipation decomposition is its potential for use as a modeling tool. Energy-based POD modes frequently serve as the basis in Galerkin projections, in which governing equations are projected on the chosen basis to produce a set of equations for the modal coefficients\citep{rempfer2000low}. The number of coefficients, and thus the number of equations to be solved, can be reduced by truncating the basis, projecting the dynamics into the subspace spanned by the truncated basis. The suitability of such ROMs relies on how well the projected dynamics approximate the full dynamics of the flow, and whether the truncation error can be compensated for by use of an additional modeling layer. One important source of truncation error lies in the multi-scale nature of turbulence in combination with the emphasis on large-scale structures assumed for energy-based POD. This is particularly an issue in the case of high-$\mathrm{Re}$ flows, where the range of scales can be considerable, and the coupling between energetic large-scale structures and small-scale dissipative structures diminishes. This generally causes an under-modeling of dissipation, for which a compensating closure model will be required.

The dissipation POD is intended as an approach to tackle this issue. As mentioned in the introduction, the energy-based POD corresponds to $L^2$ convergence of the reconstructed fields, while the dissipation-based POD enforces $H^1$ convergence, which should ensure a better convergence of the dynamics. One of the motivations for the dissipation-based POD is therefore to capture significant dynamics more efficiently. While this is only guaranteed asymptotically, we have seen this is already true at the lowest order. The dissipation-based POD identifies the C mode as the most significant modification of the LSC, which was not the case for the energy POD. \citet{soucasse2020reduced} found the $C$ mode to be necessary for the energy POD-based ROM in order to capture reorientations of the LSC. Linear stability analysis of the ROM confirmed the destabilizing influence of this mode on the LSC. 

More generally, we have established that dissipation-based POD provides a more accurate reconstruction of all energy and dissipation quantities in the boundary layers, the dynamics of  which are crucial for the Rayleigh-Bénard problem. As expected, the dissipation-based POD reconstruction of energy does not converge as fast as the energy POD in the full cell. However, Table~\ref{tab:convergence_exponents} shows that the asymptotic convergence rate is about the same for all energy and dissipation quantities reconstructed in the full cell using the dissipation based POD, suggesting optimality of the decomposition across the full range of scales.

\begin{figure}
	\includegraphics{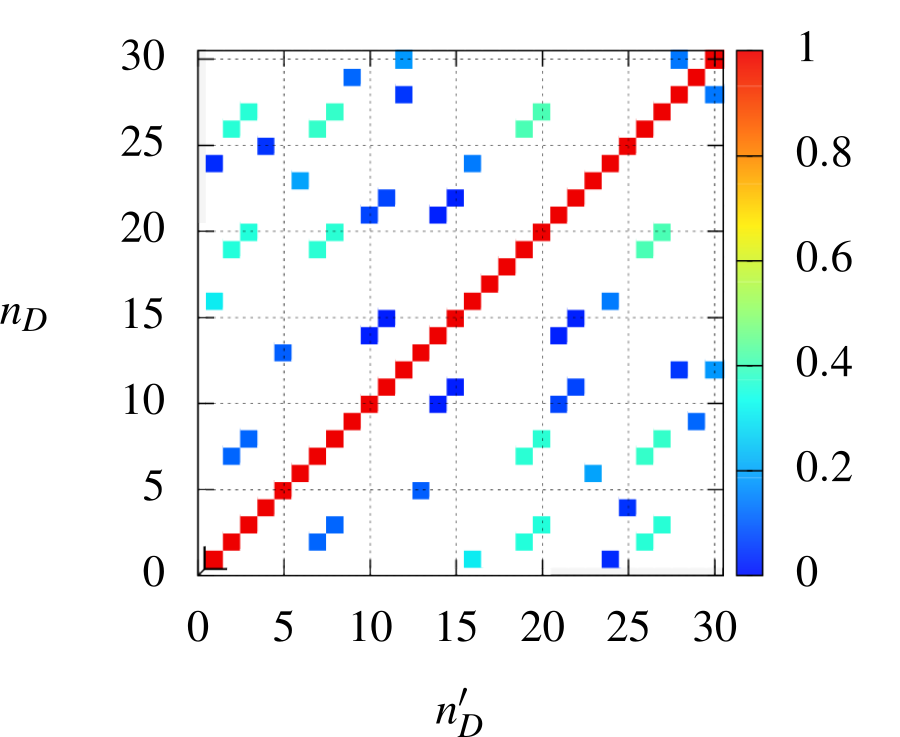}
    \caption{Non-vanishing overlap magnitudes between normalized dissipation velocity-temperature modes with $n_D,n_D'\leq 30$, computed with the energy inner product $\ip{\cdot}{\cdot}{E}$.}
    \label{fig:overlap_dissmodes}
\end{figure}

An advantage to conventional POD based Galerkin methods stems from the orthogonality of POD modes, which leads to a decoupling of the coefficient equations. The dissipation modes are not orthogonal with respect to the usual inner product, creating a potential issue when applying the Galerkin projection. However, a similar issue is faced e.g.~when working in the vorticity formulation of the Navier-Stokes equation using energy modes, as their vorticity fields are not orthogonal. \citet{rempfer1994dynamics} argued that the vorticity fields of energy eigenmodes were sufficiently close to vorticity eigenmodes that approximate orthogonality could be assumed, and showed that this assumption lead to acceptable results. In Section~\ref{sec:large_scale_organization} we observed a general trend of similarity between energy and dissipation modes, which gives some hope that at least an approximate orthogonality can be achieved for isosymmetric modes, while non-isosymmetric modes are known to be exactly orthogonal. We test this by computing the overlap matrix for the lowest 30 dissipation modes, the entries of which are shown in Figure~\ref{fig:overlap_dissmodes}. The modes are normalized such that the diagonal entries are unity. The off-diagonal part of the matrix is sparse due to the symmetry-induced constraints on mixing, and the non-zero off-diagonal entries are generally small compared to the diagonal entries ($\lesssim 0.4$). These observations give reason to believe that modal cross terms could be ignored in the present case as well, lending credibility to the feasibility of using dissipation-based velocity (or velocity-temperature) modes for ROMs. We  note that Petrov-Galerkin models allow the use of non-orthogonal bases \citep{taira2020modal}, and that data-driven learning methods such as SINDY \citep{brunton2022data} could be used to determine the ROM in the case where these approaches fail.

As seen from \eqref{eq:profile_convergence}, reconstruction of any second-order quantity can be obtained by summing the individual contributions of each mode, weighted by its eigenvalue. Figure \ref{fig:scaling} suggests that an estimate for the full quantities could be obtained from a 20-mode reconstruction, using global rescaling factors corresponding to the boundary layer and the core, with a possible matching law corresponding to the intermediate entrainment zone $0.02  \lesssim z \lesssim 0.1 $. As a first approximation, almost all the thermal dissipation and thermal energy, but only $\sim$60\% of the viscous dissipation and kinetic energy were captured by the 20-mode reconstruction in the boundary layer. Meanwhile, $\sim$50\%  of both thermal and kinetic energy were reconstructed in the core, against about respectively 15\% and 30\% for the thermal and viscous dissipation. A logarithmic interpolation between the core and the boundary layer values can be used in the entrainment region for all quantities, except for the kinetic energy reconstruction, which increased up to 75\% in the entrainment zone. We note that all results reported here have been obtained at $\mathrm{Ra}=10^7$. \citet{soucasse2021low} showed that the energy-optimized basis remained stable (with possible reordering among the modes) at least over the range $10^6 \leq \mathrm{Ra} \leq 10^8$, and proposed a $\mathrm{Ra}$-based scaling for the evolution of the dominant eigenvalues. This paves the way to an alternative POD-based approach to modeling based on determining the evolution of POD eigenvalues with the Rayleigh number. Any second-order quantity could be reconstructed over a range of Rayleigh numbers, based on a set of dissipation-optimized modes and evolution laws for the corresponding eigenvalues as well as for the velocity and temperature global scaling factors.

\section{Conclusions\label{sec:conclusion}}
We have presented a formulation of POD optimized with respect to viscous and thermal dissipation, based on replacing the usual $L^2$ norm in the POD optimization problem with an $H^1$ norm. This was motivated by the desire to improve the ability of modal decompositions to capture dynamically important structures, regardless of their energy content. The dissipation decomposition produces an orthogonal basis for the combined strain rate tensor and thermal gradient. A corresponding non-orthogonal velocity-temperature basis was produced using the extended method of snapshots, along with an energy-optimized orthogonal velocity-temperature basis using conventional snapshot POD. This allowed direct comparison between the dissipation-optimized modes and the energy-optimized POD basis. The two decomposition procedures were applied to an enriched data set based on a DNS of a Rayleigh-Bénard flow in a cubic cell, with Rayleigh number $\mathrm{Ra}=10^7$ and Prandtl number $\mathrm{Pr}=0.707$. The underlying DNS data set of $\num{1000}$ realizations was enriched based on the geometric symmetries of the convection cell, resulting in a data set of $\num{16000}$ realizations. Analyzing commutation relations for the POD and symmetry operators used for the enrichment led to a characterization of the degeneracy in the POD spectra, and the enrichment was formally shown to lead to a separation of each POD basis into distinct subspaces characterized by different sets of symmetries and anti-symmetries, constraining overlaps between modes of the different decompositions as well as the non-orthognality of dissipation modes.

The decompositions were compared in detail for the lowest 20 modes of each, for which several features were found to be shared between the two decompositions. The lowest 20 dissipation modes were found to each have a direct analog among the lowest 20 energy modes, with which both symmetry configuration and large-scale organization were shared. Much of the ordering between modes was found to be preserved between the decompositions. Both decompositions were dominated by four modes contributing to the large-scale circulation (LSC), labeled $M$, $L_{x/y}$, and $D$, which were nearly identical between the decompositions. The corner roll mode $C$, which in previous work was found to be associated  with LSC reorientations, was found to be the largest contributor to the fluctuations of the dissipation, but not those of the energy. This lends support to the hypothesis that a dissipation-based decomposition is an improvement over the energy-based decomposition in identifying dynamically significant modes.

With some exceptions dissipation modes were generally found to exhibit greater boundary layer activity and enhanced contribution to mean kinetic and thermal energy, viscous and thermal dissipation, and convective heat flux in the boundary layer compared to analogous energy modes, while showing similar or decreased contributions to mean quantities in the core of the cell. Boundary layer structures are known to play a crucial role in the dynamics of Rayleigh-Bénard flows, and the enhanced modal representation of such structures therefore suggests that the overall dynamics are more efficiently captured using the dissipation decomposition.

Differences between dissipation modes and their energy analogs tended to increase for higher modes, and particularly the roll modes $L_{x/y}^{\dagger}$ and the vertical mode $L_z$ exhibited substantial dissimilarities between the decompositions. The fact that large-scale organization was generally preserved between the decompositions was ascribed to a coupling between energetic and dissipative structures, caused by the modest Rayleigh and Reynolds numbers of the flow, at $\mathrm{Ra}=10^7$ and $\mathrm{Re} = 651$. The conservation of large-scale structures between the decompositions shows that any improvement in the resolution of boundary layers and other small-scale structure does not necessarily impact the ability of the decomposition to capture large-scale structures, suggesting instead that a wide range of scales is captured.

Modal reconstructions of global and boundary layer kinetic and thermal energy, viscous and thermal dissipation, and convective heat flux were carried out using each decomposition, and the convergence of the reconstructions was analyzed. In all cases the asymptotic convergence rates of the boundary layer reconstructions were enhanced using the dissipation decomposition compared to the energy decomposition. The asymptotic convergence rate of global reconstructions was similar for all quantities using the dissipation decomposition, while it varied considerably between energy-based reconstructions. The similar convergence rates of different quantities using dissipation modes support the hy\-po\-thes\-is that a wider range of relevant scales are captured by dissipation modes, while the improved convergence in boundary layers indicates that dynamically important boundary layer structures are more efficiently captured by these modes.

Some possible drawbacks of applying dissipation modes as a basis for Galerkin methods were also discussed. While it was argued that the non-orthogonality of dissipation-optimized velocity-temperature modes present a potential complication, the departure from orthogonality was found to be modest in the presently studied case, suggesting that the impact of this should be manageable. Based on reconstructed quantities we suggested a modeling approach based on regionally defined scaling of the flow reconstructed using the dominant eigenmodes, combined with $\mathrm{Ra}$-based scaling of dominant eigenvalues.

We have thus shown results indicating that the dissipation-based decomposition is able to capture dynamically important structures in a Rayleigh-Bénard convection flow more efficiently than conventional energy-based decompositions. The study was limited to a single configuration characterized by a moderate Rayleigh number, and it would be interesting to determine the effects of increasing the Rayleigh number. In particular, insight as to whether and how the discrepancy between the decompositions increases with the Rayleigh number would be valuable. Even at the moderate Rayleigh number considered here results suggest that the method poses a promising alternative to energy-based decompositions in the context of ROMs. Naturally, any convincing assessment of the potential of this method must revolve around the eventual implementation and evaluation of such models, which we hope will be attempted in future studies.

\begin{acknowledgments}
CMV acknowledges the support from the European Research Council: This project has received funding from the European Research Council (ERC) under the European Union's Horizon 2020 research and innovation programme (grant agreement No 803419). CMV and PJO acknowledge the support from the Poul Due Jensen Foundation: Financial support from the Poul Due Jensen Foundation (Grundfos Foundation) for this research is gratefully acknowledged.

The authors gratefully acknowledge the computational and data resources provided on the Sophia HPC Cluster at the Technical University of Denmark, DOI: 10.57940/FAFC-6M81.

This work was granted access to the HPC resources of IDRIS under the allocation 2022-A0102B00209 attributed by GENCI (Grand Équipement National de Calcul Intensif). This work was also performed using computational resources from the "Mésocentre" computing center of Université Paris-Saclay, CentraleSupélec and École Normale Supérieure Paris-Saclay supported by CNRS and Région Île-de-France (\mbox{\url{http://mesocentre.centralesupelec.fr/}}).
\end{acknowledgments}

\section*{Author declarations}
\subsection*{Conflict of interest}
The authors have no conflicts to disclose.

\subsection*{Author contributions}

\textbf{Peder J.~Olesen:} Conceptualization (equal), data curation (equal), formal analysis (equal), methodology (equal), project administration (equal), software (equal), visualization (lead), writing -- original draft preparation (lead), writing -- review and editing (lead). \textbf{Laurent Soucasse:} Conceptualization (equal), data curation (equal), formal analysis (equal), methodology (equal), project administration (equal), software (equal), writing -- original draft preparation (support), writing -- review and editing (support). \textbf{Bérengère Podvin:} Conceptualization (equal), formal analysis (equal), methodology (equal), project administration (equal), supervision (equal), writing -- original draft preparation (support), writing -- review and editing (support). \textbf{Clara M.~Velte:} Conceptualization (equal),  funding acquisition (lead), project administration (equal), supervision (equal), writing -- review and editing (support).

\subsection*{Data availability}
The data that support the findings of this study are available from the corresponding author upon reasonable request.

\section*{References}
\bibliography{biblio}

\begin{thebibliography}{51}%
\makeatletter
\providecommand \@ifxundefined [1]{%
 \@ifx{#1\undefined}
}%
\providecommand \@ifnum [1]{%
 \ifnum #1\expandafter \@firstoftwo
 \else \expandafter \@secondoftwo
 \fi
}%
\providecommand \@ifx [1]{%
 \ifx #1\expandafter \@firstoftwo
 \else \expandafter \@secondoftwo
 \fi
}%
\providecommand \natexlab [1]{#1}%
\providecommand \enquote  [1]{``#1''}%
\providecommand \bibnamefont  [1]{#1}%
\providecommand \bibfnamefont [1]{#1}%
\providecommand \citenamefont [1]{#1}%
\providecommand \href@noop [0]{\@secondoftwo}%
\providecommand \href [0]{\begingroup \@sanitize@url \@href}%
\providecommand \@href[1]{\@@startlink{#1}\@@href}%
\providecommand \@@href[1]{\endgroup#1\@@endlink}%
\providecommand \@sanitize@url [0]{\catcode `\\12\catcode `\$12\catcode
  `\&12\catcode `\#12\catcode `\^12\catcode `\_12\catcode `\%12\relax}%
\providecommand \@@startlink[1]{}%
\providecommand \@@endlink[0]{}%
\providecommand \url  [0]{\begingroup\@sanitize@url \@url }%
\providecommand \@url [1]{\endgroup\@href {#1}{\urlprefix }}%
\providecommand \urlprefix  [0]{URL }%
\providecommand \Eprint [0]{\href }%
\providecommand \doibase [0]{http://dx.doi.org/}%
\providecommand \selectlanguage [0]{\@gobble}%
\providecommand \bibinfo  [0]{\@secondoftwo}%
\providecommand \bibfield  [0]{\@secondoftwo}%
\providecommand \translation [1]{[#1]}%
\providecommand \BibitemOpen [0]{}%
\providecommand \bibitemStop [0]{}%
\providecommand \bibitemNoStop [0]{.\EOS\space}%
\providecommand \EOS [0]{\spacefactor3000\relax}%
\providecommand \BibitemShut  [1]{\csname bibitem#1\endcsname}%
\let\auto@bib@innerbib\@empty
\bibitem [{\citenamefont {Shraiman}\ and\ \citenamefont
  {Siggia}(1990)}]{shraiman1990heat}%
  \BibitemOpen
  \bibfield  {author} {\bibinfo {author} {\bibfnamefont {B.~I.}\ \bibnamefont
  {Shraiman}}\ and\ \bibinfo {author} {\bibfnamefont {E.~D.}\ \bibnamefont
  {Siggia}},\ }\bibfield  {title} {\enquote {\bibinfo {title} {Heat transport
  in high-{R}ayleigh-number convection},}\ }\href@noop {} {\bibfield  {journal}
  {\bibinfo  {journal} {Physical Review A}\ }\textbf {\bibinfo {volume} {42}},\
  \bibinfo {pages} {3650} (\bibinfo {year} {1990})}\BibitemShut {NoStop}%
\bibitem [{\citenamefont {Grossmann}\ and\ \citenamefont
  {Lohse}(2000)}]{grossmann2000scaling}%
  \BibitemOpen
  \bibfield  {author} {\bibinfo {author} {\bibfnamefont {S.}~\bibnamefont
  {Grossmann}}\ and\ \bibinfo {author} {\bibfnamefont {D.}~\bibnamefont
  {Lohse}},\ }\bibfield  {title} {\enquote {\bibinfo {title} {Scaling in
  thermal convection: a unifying theory},}\ }\href@noop {} {\bibfield
  {journal} {\bibinfo  {journal} {Journal of Fluid Mechanics}\ }\textbf
  {\bibinfo {volume} {407}},\ \bibinfo {pages} {27--56} (\bibinfo {year}
  {2000})}\BibitemShut {NoStop}%
\bibitem [{\citenamefont {Siggia}(1994)}]{siggia1994high}%
  \BibitemOpen
  \bibfield  {author} {\bibinfo {author} {\bibfnamefont {E.~D.}\ \bibnamefont
  {Siggia}},\ }\bibfield  {title} {\enquote {\bibinfo {title} {High {R}ayleigh
  number convection},}\ }\href@noop {} {\bibfield  {journal} {\bibinfo
  {journal} {Annual review of fluid mechanics}\ }\textbf {\bibinfo {volume}
  {26}},\ \bibinfo {pages} {137--168} (\bibinfo {year} {1994})}\BibitemShut
  {NoStop}%
\bibitem [{\citenamefont {Chill{\`a}}\ and\ \citenamefont
  {Schumacher}(2012)}]{chilla2012new}%
  \BibitemOpen
  \bibfield  {author} {\bibinfo {author} {\bibfnamefont {F.}~\bibnamefont
  {Chill{\`a}}}\ and\ \bibinfo {author} {\bibfnamefont {J.}~\bibnamefont
  {Schumacher}},\ }\bibfield  {title} {\enquote {\bibinfo {title} {New
  perspectives in turbulent {R}ayleigh-{B\'e}nard convection},}\ }\href@noop {}
  {\bibfield  {journal} {\bibinfo  {journal} {The European Physical Journal E}\
  }\textbf {\bibinfo {volume} {35}},\ \bibinfo {pages} {1--25} (\bibinfo {year}
  {2012})}\BibitemShut {NoStop}%
\bibitem [{\citenamefont {Scheel}\ and\ \citenamefont
  {Schumacher}(2014)}]{scheel2014local}%
  \BibitemOpen
  \bibfield  {author} {\bibinfo {author} {\bibfnamefont {J.~D.}\ \bibnamefont
  {Scheel}}\ and\ \bibinfo {author} {\bibfnamefont {J.}~\bibnamefont
  {Schumacher}},\ }\bibfield  {title} {\enquote {\bibinfo {title} {Local
  boundary layer scales in turbulent {R}ayleigh--{B\'e}nard convection},}\
  }\href@noop {} {\bibfield  {journal} {\bibinfo  {journal} {Journal of fluid
  mechanics}\ }\textbf {\bibinfo {volume} {758}},\ \bibinfo {pages} {344--373}
  (\bibinfo {year} {2014})}\BibitemShut {NoStop}%
\bibitem [{\citenamefont {Hanjali{\'c}}(2002)}]{hanjalic2002one}%
  \BibitemOpen
  \bibfield  {author} {\bibinfo {author} {\bibfnamefont {K.}~\bibnamefont
  {Hanjali{\'c}}},\ }\bibfield  {title} {\enquote {\bibinfo {title} {One-point
  closure models for buoyancy-driven turbulent flows},}\ }\href@noop {}
  {\bibfield  {journal} {\bibinfo  {journal} {Annual review of fluid
  mechanics}\ }\textbf {\bibinfo {volume} {34}},\ \bibinfo {pages} {321--347}
  (\bibinfo {year} {2002})}\BibitemShut {NoStop}%
\bibitem [{\citenamefont {Grossmann}\ and\ \citenamefont
  {Lohse}(2004)}]{grossmann2004fluctuations}%
  \BibitemOpen
  \bibfield  {author} {\bibinfo {author} {\bibfnamefont {S.}~\bibnamefont
  {Grossmann}}\ and\ \bibinfo {author} {\bibfnamefont {D.}~\bibnamefont
  {Lohse}},\ }\bibfield  {title} {\enquote {\bibinfo {title} {Fluctuations in
  turbulent {R}ayleigh--{B\'e}nard convection: the role of plumes},}\
  }\href@noop {} {\bibfield  {journal} {\bibinfo  {journal} {Physics of
  fluids}\ }\textbf {\bibinfo {volume} {16}},\ \bibinfo {pages} {4462--4472}
  (\bibinfo {year} {2004})}\BibitemShut {NoStop}%
\bibitem [{\citenamefont {Shang}\ \emph {et~al.}(2003)\citenamefont {Shang},
  \citenamefont {Qiu}, \citenamefont {Tong},\ and\ \citenamefont
  {Xia}}]{shang2003measured}%
  \BibitemOpen
  \bibfield  {author} {\bibinfo {author} {\bibfnamefont {X.-D.}\ \bibnamefont
  {Shang}}, \bibinfo {author} {\bibfnamefont {X.-L.}\ \bibnamefont {Qiu}},
  \bibinfo {author} {\bibfnamefont {P.}~\bibnamefont {Tong}}, \ and\ \bibinfo
  {author} {\bibfnamefont {K.-Q.}\ \bibnamefont {Xia}},\ }\bibfield  {title}
  {\enquote {\bibinfo {title} {Measured local heat transport in turbulent
  {R}ayleigh-{B\'e}nard convection},}\ }\href@noop {} {\bibfield  {journal}
  {\bibinfo  {journal} {Physical review letters}\ }\textbf {\bibinfo {volume}
  {90}},\ \bibinfo {pages} {074501} (\bibinfo {year} {2003})}\BibitemShut
  {NoStop}%
\bibitem [{\citenamefont {Emran}\ and\ \citenamefont
  {Schumacher}(2012)}]{emran2012conditional}%
  \BibitemOpen
  \bibfield  {author} {\bibinfo {author} {\bibfnamefont {M.~S.}\ \bibnamefont
  {Emran}}\ and\ \bibinfo {author} {\bibfnamefont {J.}~\bibnamefont
  {Schumacher}},\ }\bibfield  {title} {\enquote {\bibinfo {title} {Conditional
  statistics of thermal dissipation rate in turbulent {R}ayleigh-{B\'e}nard
  convection},}\ }\href@noop {} {\bibfield  {journal} {\bibinfo  {journal} {The
  European Physical Journal E}\ }\textbf {\bibinfo {volume} {35}},\ \bibinfo
  {pages} {1--8} (\bibinfo {year} {2012})}\BibitemShut {NoStop}%
\bibitem [{\citenamefont {Vishnu}, \citenamefont {De},\ and\ \citenamefont
  {Mishra}(2022)}]{vishnu2022statistics}%
  \BibitemOpen
  \bibfield  {author} {\bibinfo {author} {\bibfnamefont {V.~T.}\ \bibnamefont
  {Vishnu}}, \bibinfo {author} {\bibfnamefont {A.~K.}\ \bibnamefont {De}}, \
  and\ \bibinfo {author} {\bibfnamefont {P.~K.}\ \bibnamefont {Mishra}},\
  }\bibfield  {title} {\enquote {\bibinfo {title} {Statistics of thermal plumes
  and dissipation rates in turbulent rayleigh-benard convection in a cubic
  cell},}\ }\href@noop {} {\bibfield  {journal} {\bibinfo  {journal}
  {International Journal of Heat and Mass Transfer}\ }\textbf {\bibinfo
  {volume} {182}},\ \bibinfo {pages} {121995} (\bibinfo {year}
  {2022})}\BibitemShut {NoStop}%
\bibitem [{\citenamefont {Shishkina}\ and\ \citenamefont
  {Wagner}(2008)}]{shishkina2008analysis}%
  \BibitemOpen
  \bibfield  {author} {\bibinfo {author} {\bibfnamefont {O.}~\bibnamefont
  {Shishkina}}\ and\ \bibinfo {author} {\bibfnamefont {C.}~\bibnamefont
  {Wagner}},\ }\bibfield  {title} {\enquote {\bibinfo {title} {Analysis of
  sheet-like thermal plumes in turbulent {R}ayleigh--{B\'e}nard convection},}\
  }\href@noop {} {\bibfield  {journal} {\bibinfo  {journal} {Journal of Fluid
  Mechanics}\ }\textbf {\bibinfo {volume} {599}},\ \bibinfo {pages} {383--404}
  (\bibinfo {year} {2008})}\BibitemShut {NoStop}%
\bibitem [{\citenamefont {Xi}, \citenamefont {Lam},\ and\ \citenamefont
  {Xia}(2004)}]{xi2004laminar}%
  \BibitemOpen
  \bibfield  {author} {\bibinfo {author} {\bibfnamefont {H.-D.}\ \bibnamefont
  {Xi}}, \bibinfo {author} {\bibfnamefont {S.}~\bibnamefont {Lam}}, \ and\
  \bibinfo {author} {\bibfnamefont {K.-Q.}\ \bibnamefont {Xia}},\ }\bibfield
  {title} {\enquote {\bibinfo {title} {From laminar plumes to organized flows:
  the onset of large-scale circulation in turbulent thermal convection},}\
  }\href@noop {} {\bibfield  {journal} {\bibinfo  {journal} {Journal of Fluid
  Mechanics}\ }\textbf {\bibinfo {volume} {503}},\ \bibinfo {pages} {47--56}
  (\bibinfo {year} {2004})}\BibitemShut {NoStop}%
\bibitem [{\citenamefont {Hanjali{\'c}}\ and\ \citenamefont
  {Launder}(2022)}]{hanjalic2022modelling}%
  \BibitemOpen
  \bibfield  {author} {\bibinfo {author} {\bibfnamefont {K.}~\bibnamefont
  {Hanjali{\'c}}}\ and\ \bibinfo {author} {\bibfnamefont {B.}~\bibnamefont
  {Launder}},\ }\href@noop {} {\emph {\bibinfo {title} {Modelling Turbulence in
  Engineering and the Environment: Rational Alternative Routes to Closure}}}\
  (\bibinfo  {publisher} {Cambridge University Press},\ \bibinfo {year}
  {2022})\BibitemShut {NoStop}%
\bibitem [{\citenamefont {Berkooz}, \citenamefont {Holmes},\ and\ \citenamefont
  {Lumley}(1993)}]{berkooz1993proper}%
  \BibitemOpen
  \bibfield  {author} {\bibinfo {author} {\bibfnamefont {G.}~\bibnamefont
  {Berkooz}}, \bibinfo {author} {\bibfnamefont {P.}~\bibnamefont {Holmes}}, \
  and\ \bibinfo {author} {\bibfnamefont {J.~L.}\ \bibnamefont {Lumley}},\
  }\bibfield  {title} {\enquote {\bibinfo {title} {The proper orthogonal
  decomposition in the analysis of turbulent flows},}\ }\href@noop {}
  {\bibfield  {journal} {\bibinfo  {journal} {Annual review of fluid
  mechanics}\ }\textbf {\bibinfo {volume} {25}},\ \bibinfo {pages} {539--575}
  (\bibinfo {year} {1993})}\BibitemShut {NoStop}%
\bibitem [{\citenamefont {Berkooz}(1996)}]{berkooz1996turbulence}%
  \BibitemOpen
  \bibfield  {author} {\bibinfo {author} {\bibfnamefont {G.}~\bibnamefont
  {Berkooz}},\ }\href@noop {} {\emph {\bibinfo {title} {Turbulence, coherent
  structures, dynamical systems and symmetry}}}\ (\bibinfo  {publisher}
  {Cambridge University Press},\ \bibinfo {year} {1996})\BibitemShut {NoStop}%
\bibitem [{\citenamefont {Lumley}(1967)}]{lumley1967structure}%
  \BibitemOpen
  \bibfield  {author} {\bibinfo {author} {\bibfnamefont {J.~L.}\ \bibnamefont
  {Lumley}},\ }\bibfield  {title} {\enquote {\bibinfo {title} {The structure of
  inhomogeneous turbulent flows},}\ }\href@noop {} {\bibfield  {journal}
  {\bibinfo  {journal} {Atmospheric turbulence and radio wave propagation}\ ,\
  \bibinfo {pages} {166--178}} (\bibinfo {year} {1967})}\BibitemShut {NoStop}%
\bibitem [{\citenamefont {Sirovich}\ and\ \citenamefont
  {Park}(1990)}]{sirovich1990turbulent}%
  \BibitemOpen
  \bibfield  {author} {\bibinfo {author} {\bibfnamefont {L.}~\bibnamefont
  {Sirovich}}\ and\ \bibinfo {author} {\bibfnamefont {H.}~\bibnamefont
  {Park}},\ }\bibfield  {title} {\enquote {\bibinfo {title} {Turbulent thermal
  convection in a finite domain: Part {I}. {T}heory},}\ }\href@noop {}
  {\bibfield  {journal} {\bibinfo  {journal} {Physics of Fluids A: Fluid
  Dynamics}\ }\textbf {\bibinfo {volume} {2}},\ \bibinfo {pages} {1649--1658}
  (\bibinfo {year} {1990})}\BibitemShut {NoStop}%
\bibitem [{\citenamefont {Bailon-Cuba}, \citenamefont {Emran},\ and\
  \citenamefont {Schumacher}(2010)}]{bailon2010aspect}%
  \BibitemOpen
  \bibfield  {author} {\bibinfo {author} {\bibfnamefont {J.}~\bibnamefont
  {Bailon-Cuba}}, \bibinfo {author} {\bibfnamefont {M.~S.}\ \bibnamefont
  {Emran}}, \ and\ \bibinfo {author} {\bibfnamefont {J.}~\bibnamefont
  {Schumacher}},\ }\bibfield  {title} {\enquote {\bibinfo {title} {Aspect ratio
  dependence of heat transfer and large-scale flow in turbulent convection},}\
  }\href@noop {} {\bibfield  {journal} {\bibinfo  {journal} {Journal of Fluid
  Mechanics}\ }\textbf {\bibinfo {volume} {655}},\ \bibinfo {pages} {152--173}
  (\bibinfo {year} {2010})}\BibitemShut {NoStop}%
\bibitem [{\citenamefont {Podvin}\ and\ \citenamefont
  {Sergent}(2012)}]{podvin2012proper}%
  \BibitemOpen
  \bibfield  {author} {\bibinfo {author} {\bibfnamefont {B.}~\bibnamefont
  {Podvin}}\ and\ \bibinfo {author} {\bibfnamefont {A.}~\bibnamefont
  {Sergent}},\ }\bibfield  {title} {\enquote {\bibinfo {title} {Proper
  orthogonal decomposition investigation of turbulent {R}ayleigh-{B}{\'e}nard
  convection in a rectangular cavity},}\ }\href@noop {} {\bibfield  {journal}
  {\bibinfo  {journal} {Physics of Fluids}\ }\textbf {\bibinfo {volume} {24}}
  (\bibinfo {year} {2012})}\BibitemShut {NoStop}%
\bibitem [{\citenamefont {Verdoold}, \citenamefont {Tummers},\ and\
  \citenamefont {Hanjali{\'c}}(2009)}]{verdoold2009prime}%
  \BibitemOpen
  \bibfield  {author} {\bibinfo {author} {\bibfnamefont {J.}~\bibnamefont
  {Verdoold}}, \bibinfo {author} {\bibfnamefont {M.~J.}\ \bibnamefont
  {Tummers}}, \ and\ \bibinfo {author} {\bibfnamefont {K.}~\bibnamefont
  {Hanjali{\'c}}},\ }\bibfield  {title} {\enquote {\bibinfo {title} {Prime
  modes of fluid circulation in large-aspect-ratio turbulent
  {R}ayleigh-{B}{\'e}nard convection},}\ }\href@noop {} {\bibfield  {journal}
  {\bibinfo  {journal} {Physical Review E}\ }\textbf {\bibinfo {volume} {80}},\
  \bibinfo {pages} {037301} (\bibinfo {year} {2009})}\BibitemShut {NoStop}%
\bibitem [{\citenamefont {Podvin}\ and\ \citenamefont
  {Sergent}(2015)}]{podvin2015large}%
  \BibitemOpen
  \bibfield  {author} {\bibinfo {author} {\bibfnamefont {B.}~\bibnamefont
  {Podvin}}\ and\ \bibinfo {author} {\bibfnamefont {A.}~\bibnamefont
  {Sergent}},\ }\bibfield  {title} {\enquote {\bibinfo {title} {A large-scale
  investigation of wind reversal in a square {Rayleigh--B\'enard} cell},}\
  }\href@noop {} {\bibfield  {journal} {\bibinfo  {journal} {Journal of Fluid
  Mechanics}\ }\textbf {\bibinfo {volume} {766}},\ \bibinfo {pages} {172--201}
  (\bibinfo {year} {2015})}\BibitemShut {NoStop}%
\bibitem [{\citenamefont {Soucasse}\ \emph {et~al.}(2019)\citenamefont
  {Soucasse}, \citenamefont {Podvin}, \citenamefont {Rivi{\`e}re},\ and\
  \citenamefont {Soufiani}}]{soucasse2019proper}%
  \BibitemOpen
  \bibfield  {author} {\bibinfo {author} {\bibfnamefont {L.}~\bibnamefont
  {Soucasse}}, \bibinfo {author} {\bibfnamefont {B.}~\bibnamefont {Podvin}},
  \bibinfo {author} {\bibfnamefont {P.}~\bibnamefont {Rivi{\`e}re}}, \ and\
  \bibinfo {author} {\bibfnamefont {A.}~\bibnamefont {Soufiani}},\ }\bibfield
  {title} {\enquote {\bibinfo {title} {Proper orthogonal decomposition analysis
  and modelling of large-scale flow reorientations in a cubic
  {R}ayleigh--{B\'e}nard cell},}\ }\href@noop {} {\bibfield  {journal}
  {\bibinfo  {journal} {Journal of Fluid Mechanics}\ }\textbf {\bibinfo
  {volume} {881}},\ \bibinfo {pages} {23--50} (\bibinfo {year}
  {2019})}\BibitemShut {NoStop}%
\bibitem [{\citenamefont {Bergmann}, \citenamefont {Bruneau},\ and\
  \citenamefont {Iollo}(2009)}]{bergmann2009enablers}%
  \BibitemOpen
  \bibfield  {author} {\bibinfo {author} {\bibfnamefont {M.}~\bibnamefont
  {Bergmann}}, \bibinfo {author} {\bibfnamefont {C.-H.}\ \bibnamefont
  {Bruneau}}, \ and\ \bibinfo {author} {\bibfnamefont {A.}~\bibnamefont
  {Iollo}},\ }\bibfield  {title} {\enquote {\bibinfo {title} {Enablers for
  robust {POD} models},}\ }\href@noop {} {\bibfield  {journal} {\bibinfo
  {journal} {Journal of Computational Physics}\ }\textbf {\bibinfo {volume}
  {228}},\ \bibinfo {pages} {516--538} (\bibinfo {year} {2009})}\BibitemShut
  {NoStop}%
\bibitem [{\citenamefont {Aubry}, \citenamefont {Lian},\ and\ \citenamefont
  {Titi}(1993)}]{aubry1993preserving}%
  \BibitemOpen
  \bibfield  {author} {\bibinfo {author} {\bibfnamefont {N.}~\bibnamefont
  {Aubry}}, \bibinfo {author} {\bibfnamefont {W.-Y.}\ \bibnamefont {Lian}}, \
  and\ \bibinfo {author} {\bibfnamefont {E.~S.}\ \bibnamefont {Titi}},\
  }\bibfield  {title} {\enquote {\bibinfo {title} {Preserving symmetries in the
  proper orthogonal decomposition},}\ }\href@noop {} {\bibfield  {journal}
  {\bibinfo  {journal} {SIAM Journal on Scientific Computing}\ }\textbf
  {\bibinfo {volume} {14}},\ \bibinfo {pages} {483--505} (\bibinfo {year}
  {1993})}\BibitemShut {NoStop}%
\bibitem [{\citenamefont {Sengupta}\ and\ \citenamefont
  {Dey}(2004)}]{sengupta2004proper}%
  \BibitemOpen
  \bibfield  {author} {\bibinfo {author} {\bibfnamefont {T.~K.}\ \bibnamefont
  {Sengupta}}\ and\ \bibinfo {author} {\bibfnamefont {S.}~\bibnamefont {Dey}},\
  }\bibfield  {title} {\enquote {\bibinfo {title} {Proper orthogonal
  decomposition of direct numerical simulation data of by-pass transition},}\
  }\href@noop {} {\bibfield  {journal} {\bibinfo  {journal} {Computers \&
  structures}\ }\textbf {\bibinfo {volume} {82}},\ \bibinfo {pages}
  {2693--2703} (\bibinfo {year} {2004})}\BibitemShut {NoStop}%
\bibitem [{\citenamefont {Lee}\ and\ \citenamefont
  {Dowell}(2020)}]{lee2020improving}%
  \BibitemOpen
  \bibfield  {author} {\bibinfo {author} {\bibfnamefont {M.~W.}\ \bibnamefont
  {Lee}}\ and\ \bibinfo {author} {\bibfnamefont {E.~H.}\ \bibnamefont
  {Dowell}},\ }\bibfield  {title} {\enquote {\bibinfo {title} {Improving the
  predictable accuracy of fluid galerkin reduced-order models using two {POD}
  bases},}\ }\href@noop {} {\bibfield  {journal} {\bibinfo  {journal}
  {Nonlinear Dynamics}\ }\textbf {\bibinfo {volume} {101}},\ \bibinfo {pages}
  {1457--1471} (\bibinfo {year} {2020})}\BibitemShut {NoStop}%
\bibitem [{\citenamefont {Schi{\o}dt}\ \emph {et~al.}(2022)\citenamefont
  {Schi{\o}dt}, \citenamefont {Hod{\v{z}}i{\'c}}, \citenamefont {Evrard},
  \citenamefont {Hausmann}, \citenamefont {Van~Wachem},\ and\ \citenamefont
  {Velte}}]{schiodt2022characterizing}%
  \BibitemOpen
  \bibfield  {author} {\bibinfo {author} {\bibfnamefont {M.}~\bibnamefont
  {Schi{\o}dt}}, \bibinfo {author} {\bibfnamefont {A.}~\bibnamefont
  {Hod{\v{z}}i{\'c}}}, \bibinfo {author} {\bibfnamefont {F.}~\bibnamefont
  {Evrard}}, \bibinfo {author} {\bibfnamefont {M.}~\bibnamefont {Hausmann}},
  \bibinfo {author} {\bibfnamefont {B.}~\bibnamefont {Van~Wachem}}, \ and\
  \bibinfo {author} {\bibfnamefont {C.~M.}\ \bibnamefont {Velte}},\ }\bibfield
  {title} {\enquote {\bibinfo {title} {Characterizing {L}agrangian particle
  dynamics in decaying homogeneous isotropic turbulence using proper orthogonal
  decomposition},}\ }\href@noop {} {\bibfield  {journal} {\bibinfo  {journal}
  {Physics of Fluids}\ }\textbf {\bibinfo {volume} {34}} (\bibinfo {year}
  {2022})}\BibitemShut {NoStop}%
\bibitem [{\citenamefont {Bor{\'e}e}(2003)}]{boree2003extended}%
  \BibitemOpen
  \bibfield  {author} {\bibinfo {author} {\bibfnamefont {J.}~\bibnamefont
  {Bor{\'e}e}},\ }\bibfield  {title} {\enquote {\bibinfo {title} {Extended
  proper orthogonal decomposition: a tool to analyse correlated events in
  turbulent flows},}\ }\href@noop {} {\bibfield  {journal} {\bibinfo  {journal}
  {Experiments in fluids}\ }\textbf {\bibinfo {volume} {35}},\ \bibinfo {pages}
  {188--192} (\bibinfo {year} {2003})}\BibitemShut {NoStop}%
\bibitem [{\citenamefont {Olesen}\ \emph {et~al.}(2023)\citenamefont {Olesen},
  \citenamefont {Hod{\v{z}}i{\'c}}, \citenamefont {Andersen}, \citenamefont
  {S{\o}rensen},\ and\ \citenamefont {Velte}}]{olesen2023dissipation}%
  \BibitemOpen
  \bibfield  {author} {\bibinfo {author} {\bibfnamefont {P.~J.}\ \bibnamefont
  {Olesen}}, \bibinfo {author} {\bibfnamefont {A.}~\bibnamefont
  {Hod{\v{z}}i{\'c}}}, \bibinfo {author} {\bibfnamefont {S.~J.}\ \bibnamefont
  {Andersen}}, \bibinfo {author} {\bibfnamefont {N.~N.}\ \bibnamefont
  {S{\o}rensen}}, \ and\ \bibinfo {author} {\bibfnamefont {C.~M.}\ \bibnamefont
  {Velte}},\ }\bibfield  {title} {\enquote {\bibinfo {title}
  {Dissipation-optimized proper orthogonal decomposition},}\ }\href@noop {}
  {\bibfield  {journal} {\bibinfo  {journal} {Physics of Fluids}\ }\textbf
  {\bibinfo {volume} {35}} (\bibinfo {year} {2023})}\BibitemShut {NoStop}%
\bibitem [{\citenamefont {Bakewell~Jr}\ and\ \citenamefont
  {Lumley}(1967)}]{bakewell1967viscous}%
  \BibitemOpen
  \bibfield  {author} {\bibinfo {author} {\bibfnamefont {H.~P.}\ \bibnamefont
  {Bakewell~Jr}}\ and\ \bibinfo {author} {\bibfnamefont {J.~L.}\ \bibnamefont
  {Lumley}},\ }\bibfield  {title} {\enquote {\bibinfo {title} {Viscous sublayer
  and adjacent wall region in turbulent pipe flow},}\ }\href@noop {} {\bibfield
   {journal} {\bibinfo  {journal} {The Physics of Fluids}\ }\textbf {\bibinfo
  {volume} {10}},\ \bibinfo {pages} {1880--1889} (\bibinfo {year}
  {1967})}\BibitemShut {NoStop}%
\bibitem [{\citenamefont {Herzog}(1986)}]{herzog1986large}%
  \BibitemOpen
  \bibfield  {author} {\bibinfo {author} {\bibfnamefont {S.}~\bibnamefont
  {Herzog}},\ }\href@noop {} {\emph {\bibinfo {title} {The large scale
  structure in the near-wall region of turbulent pipe flow}}}\ (\bibinfo
  {publisher} {Cornell University},\ \bibinfo {year} {1986})\BibitemShut
  {NoStop}%
\bibitem [{\citenamefont {Aubry}\ \emph {et~al.}(1988)\citenamefont {Aubry},
  \citenamefont {Holmes}, \citenamefont {Lumley},\ and\ \citenamefont
  {Stone}}]{aubry1988dynamics}%
  \BibitemOpen
  \bibfield  {author} {\bibinfo {author} {\bibfnamefont {N.}~\bibnamefont
  {Aubry}}, \bibinfo {author} {\bibfnamefont {P.}~\bibnamefont {Holmes}},
  \bibinfo {author} {\bibfnamefont {J.~L.}\ \bibnamefont {Lumley}}, \ and\
  \bibinfo {author} {\bibfnamefont {E.}~\bibnamefont {Stone}},\ }\bibfield
  {title} {\enquote {\bibinfo {title} {The dynamics of coherent structures in
  the wall region of a turbulent boundary layer},}\ }\href@noop {} {\bibfield
  {journal} {\bibinfo  {journal} {Journal of fluid Mechanics}\ }\textbf
  {\bibinfo {volume} {192}},\ \bibinfo {pages} {115--173} (\bibinfo {year}
  {1988})}\BibitemShut {NoStop}%
\bibitem [{\citenamefont {Deane}\ \emph {et~al.}(1991)\citenamefont {Deane},
  \citenamefont {Kevrekidis}, \citenamefont {Karniadakis},\ and\ \citenamefont
  {Orszag}}]{deane1991low}%
  \BibitemOpen
  \bibfield  {author} {\bibinfo {author} {\bibfnamefont {A.}~\bibnamefont
  {Deane}}, \bibinfo {author} {\bibfnamefont {I.}~\bibnamefont {Kevrekidis}},
  \bibinfo {author} {\bibfnamefont {G.~E.}\ \bibnamefont {Karniadakis}}, \ and\
  \bibinfo {author} {\bibfnamefont {S.}~\bibnamefont {Orszag}},\ }\bibfield
  {title} {\enquote {\bibinfo {title} {Low-dimensional models for complex
  geometry flows: Application to grooved channels and circular cylinders},}\
  }\href@noop {} {\bibfield  {journal} {\bibinfo  {journal} {Physics of Fluids
  A: Fluid Dynamics}\ }\textbf {\bibinfo {volume} {3}},\ \bibinfo {pages}
  {2337--2354} (\bibinfo {year} {1991})}\BibitemShut {NoStop}%
\bibitem [{\citenamefont {Ravindran}(2000)}]{ravindran2000reduced}%
  \BibitemOpen
  \bibfield  {author} {\bibinfo {author} {\bibfnamefont {S.~S.}\ \bibnamefont
  {Ravindran}},\ }\bibfield  {title} {\enquote {\bibinfo {title} {A
  reduced-order approach for optimal control of fluids using proper orthogonal
  decomposition},}\ }\href@noop {} {\bibfield  {journal} {\bibinfo  {journal}
  {International journal for numerical methods in fluids}\ }\textbf {\bibinfo
  {volume} {34}},\ \bibinfo {pages} {425--448} (\bibinfo {year}
  {2000})}\BibitemShut {NoStop}%
\bibitem [{\citenamefont {Ly}\ and\ \citenamefont
  {Tran}(2001)}]{ly2001modeling}%
  \BibitemOpen
  \bibfield  {author} {\bibinfo {author} {\bibfnamefont {H.~V.}\ \bibnamefont
  {Ly}}\ and\ \bibinfo {author} {\bibfnamefont {H.~T.}\ \bibnamefont {Tran}},\
  }\bibfield  {title} {\enquote {\bibinfo {title} {Modeling and control of
  physical processes using proper orthogonal decomposition},}\ }\href@noop {}
  {\bibfield  {journal} {\bibinfo  {journal} {Mathematical and computer
  modelling}\ }\textbf {\bibinfo {volume} {33}},\ \bibinfo {pages} {223--236}
  (\bibinfo {year} {2001})}\BibitemShut {NoStop}%
\bibitem [{\citenamefont {Sirovich}(1987)}]{sirovich1987turbulence}%
  \BibitemOpen
  \bibfield  {author} {\bibinfo {author} {\bibfnamefont {L.}~\bibnamefont
  {Sirovich}},\ }\bibfield  {title} {\enquote {\bibinfo {title} {Turbulence and
  the dynamics of coherent structures. {I}. {C}oherent structures},}\
  }\href@noop {} {\bibfield  {journal} {\bibinfo  {journal} {Quarterly of
  applied mathematics}\ }\textbf {\bibinfo {volume} {45}},\ \bibinfo {pages}
  {561--571} (\bibinfo {year} {1987})}\BibitemShut {NoStop}%
\bibitem [{\citenamefont {Zhang}\ \emph {et~al.}(2023)\citenamefont {Zhang},
  \citenamefont {Hod{\v{z}}i{\'c}}, \citenamefont {Evrard}, \citenamefont
  {Van~Wachem},\ and\ \citenamefont {Velte}}]{zhang2023phase}%
  \BibitemOpen
  \bibfield  {author} {\bibinfo {author} {\bibfnamefont {Y.}~\bibnamefont
  {Zhang}}, \bibinfo {author} {\bibfnamefont {A.}~\bibnamefont
  {Hod{\v{z}}i{\'c}}}, \bibinfo {author} {\bibfnamefont {F.}~\bibnamefont
  {Evrard}}, \bibinfo {author} {\bibfnamefont {B.}~\bibnamefont {Van~Wachem}},
  \ and\ \bibinfo {author} {\bibfnamefont {C.~M.}\ \bibnamefont {Velte}},\
  }\bibfield  {title} {\enquote {\bibinfo {title} {Phase proper orthogonal
  decomposition of non-stationary turbulent flow},}\ }\href@noop {} {\bibfield
  {journal} {\bibinfo  {journal} {Physics of Fluids}\ }\textbf {\bibinfo
  {volume} {35}},\ \bibinfo {pages} {045109} (\bibinfo {year}
  {2023})}\BibitemShut {NoStop}%
\bibitem [{\citenamefont {Towne}, \citenamefont {Schmidt},\ and\ \citenamefont
  {Colonius}(2018)}]{towne2018spectral}%
  \BibitemOpen
  \bibfield  {author} {\bibinfo {author} {\bibfnamefont {A.}~\bibnamefont
  {Towne}}, \bibinfo {author} {\bibfnamefont {O.~T.}\ \bibnamefont {Schmidt}},
  \ and\ \bibinfo {author} {\bibfnamefont {T.}~\bibnamefont {Colonius}},\
  }\bibfield  {title} {\enquote {\bibinfo {title} {Spectral proper orthogonal
  decomposition and its relationship to dynamic mode decomposition and
  resolvent analysis},}\ }\href@noop {} {\bibfield  {journal} {\bibinfo
  {journal} {Journal of Fluid Mechanics}\ }\textbf {\bibinfo {volume} {847}},\
  \bibinfo {pages} {821--867} (\bibinfo {year} {2018})}\BibitemShut {NoStop}%
\bibitem [{\citenamefont {Sieber}, \citenamefont {Paschereit},\ and\
  \citenamefont {Oberleithner}(2016)}]{sieber2016spectral}%
  \BibitemOpen
  \bibfield  {author} {\bibinfo {author} {\bibfnamefont {M.}~\bibnamefont
  {Sieber}}, \bibinfo {author} {\bibfnamefont {C.~O.}\ \bibnamefont
  {Paschereit}}, \ and\ \bibinfo {author} {\bibfnamefont {K.}~\bibnamefont
  {Oberleithner}},\ }\bibfield  {title} {\enquote {\bibinfo {title} {Spectral
  proper orthogonal decomposition},}\ }\href@noop {} {\bibfield  {journal}
  {\bibinfo  {journal} {Journal of Fluid Mechanics}\ }\textbf {\bibinfo
  {volume} {792}},\ \bibinfo {pages} {798--828} (\bibinfo {year}
  {2016})}\BibitemShut {NoStop}%
\bibitem [{\citenamefont {Hod{\v{z}}i{\'c}}, \citenamefont {Olesen},\ and\
  \citenamefont {Velte}(2022)}]{hodvzic2022discrepancies}%
  \BibitemOpen
  \bibfield  {author} {\bibinfo {author} {\bibfnamefont {A.}~\bibnamefont
  {Hod{\v{z}}i{\'c}}}, \bibinfo {author} {\bibfnamefont {P.~J.}\ \bibnamefont
  {Olesen}}, \ and\ \bibinfo {author} {\bibfnamefont {C.~M.}\ \bibnamefont
  {Velte}},\ }\bibfield  {title} {\enquote {\bibinfo {title} {On the
  discrepancies between {POD} and {F}ourier modes on aperiodic domains},}\
  }\href@noop {} {\bibfield  {journal} {\bibinfo  {journal} {arXiv preprint
  arXiv:2207.02550}\ } (\bibinfo {year} {2022})}\BibitemShut {NoStop}%
\bibitem [{\citenamefont {Xin}\ and\ \citenamefont {{Le
  Qu{\'e}r{\'e}}}(2002)}]{xin02}%
  \BibitemOpen
  \bibfield  {author} {\bibinfo {author} {\bibfnamefont {S.}~\bibnamefont
  {Xin}}\ and\ \bibinfo {author} {\bibfnamefont {P.}~\bibnamefont {{Le
  Qu{\'e}r{\'e}}}},\ }\bibfield  {title} {\enquote {\bibinfo {title} {{An
  extended Chebyshev pseudo-spectral benchmark for the 8:1 differentially
  heated cavity}},}\ }\href@noop {} {\bibfield  {journal} {\bibinfo  {journal}
  {Numerical Metholds in Fluids}\ }\textbf {\bibinfo {volume} {40}},\ \bibinfo
  {pages} {981--998} (\bibinfo {year} {2002})}\BibitemShut {NoStop}%
\bibitem [{\citenamefont {Xin}, \citenamefont {Chergui},\ and\ \citenamefont
  {{Le Qu\'er\'e}}(2008)}]{xin-PCFD08}%
  \BibitemOpen
  \bibfield  {author} {\bibinfo {author} {\bibfnamefont {S.}~\bibnamefont
  {Xin}}, \bibinfo {author} {\bibfnamefont {J.}~\bibnamefont {Chergui}}, \ and\
  \bibinfo {author} {\bibfnamefont {P.}~\bibnamefont {{Le Qu\'er\'e}}},\
  }\bibfield  {title} {\enquote {\bibinfo {title} {3{D} spectral parallel
  multi-domain computing for natural convection flows},}\ }in\ \href@noop {}
  {\emph {\bibinfo {booktitle} {Parallel Computational Fluid Dynamics, Lecture
  Notes in Computational Science and Engineering book series}}},\ Vol.~\bibinfo
  {volume} {74},\ \bibinfo {editor} {edited by\ \bibinfo {editor} {\bibnamefont
  {Springer}}}\ (\bibinfo {year} {2008})\ pp.\ \bibinfo {pages}
  {163--171}\BibitemShut {NoStop}%
\bibitem [{\citenamefont {Shishkina}\ \emph {et~al.}(2010)\citenamefont
  {Shishkina}, \citenamefont {Stevens}, \citenamefont {Grossmann},\ and\
  \citenamefont {Lohse}}]{shishkina2010boundary}%
  \BibitemOpen
  \bibfield  {author} {\bibinfo {author} {\bibfnamefont {O.}~\bibnamefont
  {Shishkina}}, \bibinfo {author} {\bibfnamefont {R.~J.}\ \bibnamefont
  {Stevens}}, \bibinfo {author} {\bibfnamefont {S.}~\bibnamefont {Grossmann}},
  \ and\ \bibinfo {author} {\bibfnamefont {D.}~\bibnamefont {Lohse}},\
  }\bibfield  {title} {\enquote {\bibinfo {title} {Boundary layer structure in
  turbulent thermal convection and its consequences for the required numerical
  resolution},}\ }\href@noop {} {\bibfield  {journal} {\bibinfo  {journal} {New
  Journal of Physics}\ }\textbf {\bibinfo {volume} {12}},\ \bibinfo {pages}
  {075022} (\bibinfo {year} {2010})}\BibitemShut {NoStop}%
\bibitem [{\citenamefont {Delort-Laval}\ \emph {et~al.}(2022)\citenamefont
  {Delort-Laval}, \citenamefont {Soucasse}, \citenamefont {Rivière},\ and\
  \citenamefont {Soufiani}}]{delort2022rayleigh}%
  \BibitemOpen
  \bibfield  {author} {\bibinfo {author} {\bibfnamefont {M.}~\bibnamefont
  {Delort-Laval}}, \bibinfo {author} {\bibfnamefont {L.}~\bibnamefont
  {Soucasse}}, \bibinfo {author} {\bibfnamefont {P.}~\bibnamefont {Rivière}},
  \ and\ \bibinfo {author} {\bibfnamefont {A.}~\bibnamefont {Soufiani}},\
  }\bibfield  {title} {\enquote {\bibinfo {title} {Rayleigh--{B\'e}nard
  convection in a cubic cell under the effects of gas radiation up to
  $\mathrm{Ra}=10^9$},}\ }\href@noop {} {\bibfield  {journal} {\bibinfo
  {journal} {International Journal of Heat and Mass Transfer}\ }\textbf
  {\bibinfo {volume} {187}},\ \bibinfo {pages} {122453} (\bibinfo {year}
  {2022})}\BibitemShut {NoStop}%
\bibitem [{\citenamefont {Puigjaner}\ \emph {et~al.}(2008)\citenamefont
  {Puigjaner}, \citenamefont {Herrero}, \citenamefont {Simo},\ and\
  \citenamefont {Giralt}}]{puigjaner2008bifurcation}%
  \BibitemOpen
  \bibfield  {author} {\bibinfo {author} {\bibfnamefont {D.}~\bibnamefont
  {Puigjaner}}, \bibinfo {author} {\bibfnamefont {J.}~\bibnamefont {Herrero}},
  \bibinfo {author} {\bibfnamefont {C.}~\bibnamefont {Simo}}, \ and\ \bibinfo
  {author} {\bibfnamefont {F.}~\bibnamefont {Giralt}},\ }\bibfield  {title}
  {\enquote {\bibinfo {title} {Bifurcation analysis of steady
  {R}ayleigh--{B\'e}nard convection in a cubical cavity with conducting
  sidewalls},}\ }\href@noop {} {\bibfield  {journal} {\bibinfo  {journal}
  {Journal of Fluid Mechanics}\ }\textbf {\bibinfo {volume} {598}},\ \bibinfo
  {pages} {393--427} (\bibinfo {year} {2008})}\BibitemShut {NoStop}%
\bibitem [{\citenamefont {Soucasse}\ \emph {et~al.}(2021)\citenamefont
  {Soucasse}, \citenamefont {Podvin}, \citenamefont {Rivi{\`e}re},\ and\
  \citenamefont {Soufiani}}]{soucasse2021low}%
  \BibitemOpen
  \bibfield  {author} {\bibinfo {author} {\bibfnamefont {L.}~\bibnamefont
  {Soucasse}}, \bibinfo {author} {\bibfnamefont {B.}~\bibnamefont {Podvin}},
  \bibinfo {author} {\bibfnamefont {P.}~\bibnamefont {Rivi{\`e}re}}, \ and\
  \bibinfo {author} {\bibfnamefont {A.}~\bibnamefont {Soufiani}},\ }\bibfield
  {title} {\enquote {\bibinfo {title} {Low-order models for predicting
  radiative transfer effects on {Rayleigh--B\'enard} convection in a cubic cell
  at different {Rayleigh} numbers},}\ }\href@noop {} {\bibfield  {journal}
  {\bibinfo  {journal} {Journal of Fluid Mechanics}\ }\textbf {\bibinfo
  {volume} {917}},\ \bibinfo {pages} {A5} (\bibinfo {year} {2021})}\BibitemShut
  {NoStop}%
\bibitem [{\citenamefont {Soucasse}\ \emph {et~al.}(2020)\citenamefont
  {Soucasse}, \citenamefont {Podvin}, \citenamefont {Rivi{\`e}re},\ and\
  \citenamefont {Soufiani}}]{soucasse2020reduced}%
  \BibitemOpen
  \bibfield  {author} {\bibinfo {author} {\bibfnamefont {L.}~\bibnamefont
  {Soucasse}}, \bibinfo {author} {\bibfnamefont {B.}~\bibnamefont {Podvin}},
  \bibinfo {author} {\bibfnamefont {P.}~\bibnamefont {Rivi{\`e}re}}, \ and\
  \bibinfo {author} {\bibfnamefont {A.}~\bibnamefont {Soufiani}},\ }\bibfield
  {title} {\enquote {\bibinfo {title} {Reduced-order modelling of radiative
  transfer effects on {R}ayleigh--{B\'e}nard convection in a cubic cell},}\
  }\href@noop {} {\bibfield  {journal} {\bibinfo  {journal} {Journal of Fluid
  Mechanics}\ }\textbf {\bibinfo {volume} {898}},\ \bibinfo {pages} {A2}
  (\bibinfo {year} {2020})}\BibitemShut {NoStop}%
\bibitem [{\citenamefont {Rempfer}(2000)}]{rempfer2000low}%
  \BibitemOpen
  \bibfield  {author} {\bibinfo {author} {\bibfnamefont {D.}~\bibnamefont
  {Rempfer}},\ }\bibfield  {title} {\enquote {\bibinfo {title} {On
  low-dimensional {G}alerkin models for fluid flow},}\ }\href@noop {}
  {\bibfield  {journal} {\bibinfo  {journal} {Theoretical and Computational
  Fluid Dynamics}\ }\textbf {\bibinfo {volume} {14}},\ \bibinfo {pages}
  {75--88} (\bibinfo {year} {2000})}\BibitemShut {NoStop}%
\bibitem [{\citenamefont {Rempfer}\ and\ \citenamefont
  {Fasel}(1994)}]{rempfer1994dynamics}%
  \BibitemOpen
  \bibfield  {author} {\bibinfo {author} {\bibfnamefont {D.}~\bibnamefont
  {Rempfer}}\ and\ \bibinfo {author} {\bibfnamefont {H.~F.}\ \bibnamefont
  {Fasel}},\ }\bibfield  {title} {\enquote {\bibinfo {title} {Dynamics of
  three-dimensional coherent structures in a flat-plate boundary layer},}\
  }\href@noop {} {\bibfield  {journal} {\bibinfo  {journal} {Journal of Fluid
  Mechanics}\ }\textbf {\bibinfo {volume} {275}},\ \bibinfo {pages} {257--283}
  (\bibinfo {year} {1994})}\BibitemShut {NoStop}%
\bibitem [{\citenamefont {Taira}\ \emph {et~al.}(2020)\citenamefont {Taira},
  \citenamefont {Hemati}, \citenamefont {Brunton}, \citenamefont {Sun},
  \citenamefont {Duraisamy}, \citenamefont {Bagheri}, \citenamefont {Dawson},\
  and\ \citenamefont {Yeh}}]{taira2020modal}%
  \BibitemOpen
  \bibfield  {author} {\bibinfo {author} {\bibfnamefont {K.}~\bibnamefont
  {Taira}}, \bibinfo {author} {\bibfnamefont {M.~S.}\ \bibnamefont {Hemati}},
  \bibinfo {author} {\bibfnamefont {S.~L.}\ \bibnamefont {Brunton}}, \bibinfo
  {author} {\bibfnamefont {Y.}~\bibnamefont {Sun}}, \bibinfo {author}
  {\bibfnamefont {K.}~\bibnamefont {Duraisamy}}, \bibinfo {author}
  {\bibfnamefont {S.}~\bibnamefont {Bagheri}}, \bibinfo {author} {\bibfnamefont
  {S.~T.}\ \bibnamefont {Dawson}}, \ and\ \bibinfo {author} {\bibfnamefont
  {C.-A.}\ \bibnamefont {Yeh}},\ }\bibfield  {title} {\enquote {\bibinfo
  {title} {Modal analysis of fluid flows: Applications and outlook},}\
  }\href@noop {} {\bibfield  {journal} {\bibinfo  {journal} {AIAA journal}\
  }\textbf {\bibinfo {volume} {58}},\ \bibinfo {pages} {998--1022} (\bibinfo
  {year} {2020})}\BibitemShut {NoStop}%
\bibitem [{\citenamefont {Brunton}\ and\ \citenamefont
  {Kutz}(2022)}]{brunton2022data}%
  \BibitemOpen
  \bibfield  {author} {\bibinfo {author} {\bibfnamefont {S.~L.}\ \bibnamefont
  {Brunton}}\ and\ \bibinfo {author} {\bibfnamefont {J.~N.}\ \bibnamefont
  {Kutz}},\ }\href@noop {} {\emph {\bibinfo {title} {Data-driven science and
  engineering: Machine learning, dynamical systems, and control}}}\ (\bibinfo
  {publisher} {Cambridge University Press},\ \bibinfo {year}
  {2022})\BibitemShut {NoStop}%
\end{thebibliography}%

\end{document}